\documentclass[10pt,journal]{IEEEtran}
\usepackage{amssymb}
\usepackage{fontawesome}
\usepackage{algorithmicx}
\usepackage{array}
\usepackage{booktabs}
\usepackage {cite}
\usepackage{cuted}
\usepackage{bbm}
\usepackage{amsmath}
\usepackage{color}
\usepackage{enumerate}
\usepackage{graphicx}
\usepackage{epstopdf} 
\usepackage{balance}
\usepackage{ifpdf}
\usepackage{pgfplots}
\usepackage{algorithm}
\usepackage[]{algpseudocode}
\usepackage{textcomp}
\usepackage{mathtools}
\usepackage{epsfig}
\usepackage{lipsum}
\usepackage{mathtools}
\usepackage{cuted}
\usepackage{pstool}
\usepackage{psfrag}
\usepackage{comment}
\usepackage{tabularx}
\usepackage{caption}
\usepackage{subcaption}

\newtheorem{theorem}{Theorem}

\mathchardef\Re="023C
\mathchardef\Im="023D

\newtheorem{remark}{Remark}
\newtheorem{definition}{Definition}

\def\BibTeX{{\rm B\kern-.05em{\sc i\kern-.025em b}\kern-.08em
    T\kern-.1667em\lower.7ex\hbox{E}\kern-.125emX}}

\DeclareCaptionLabelSeparator{periodspace}{.\quad}
\usepackage{algpseudocode,algorithm}
\makeatletter
\newcommand{\algmargin}{\the\ALG@thistlm}
 
\DeclareMathOperator*{\argmin}{arg\,min}

\algnewcommand{\parState}[1]{\State%
  \parbox[t]{\dimexpr\linewidth-\algmargin}{\strut #1\strut}}

\pgfplotsset{compat=1.15}

\begin{document}



\title{
Near-Field Localization with RIS via Two-Dimensional Signal Path Classification
\thanks{This work was supported by the National Research Foundation of Korea (NRF) grant funded by the Korea government (MSIT) (No. NRF-2023R1A2C3002890).
J. Kang and S. Kim are with the Department of Electronics and Computer Engineering, Hanyang University, Seoul, Korea (email: \{rkdwjddhks77, remero\}@hanyang.ac.kr).  S.-W. Ko is with the Department of Smart Mobility Engineering, Inha University, Incheon, Korea (e-mail: swko@inha.ac.kr). S.-W. Ko and S. Kim are the corresponding authors.
This paper was presented in part at the IEEE GLOBECOM 2022 \cite{3532322}.
}}
\author{Jeongwan~Kang,~\IEEEmembership{Graduate~Student~Member,~IEEE,}
		Seung-Woo~Ko,~\IEEEmembership{Senior~Member,~IEEE,}
		and~Sunwoo~Kim,~\IEEEmembership{Senior~Member,~IEEE}
		}

\date{Sep 2022}
\maketitle
\begin{abstract}
In this paper, we propose \emph{two-dimensional signal path classification} (2D-SPC) for \emph{reconfigurable intelligent surface} (RIS)-assisted \emph{near-field} (NF) localization. In the NF regime, multiple RIS-driven \emph{signal paths} (SPs) can contribute to precise localization if these are decomposable and the reflected locations on the RIS are known, referred to as  \emph{SP decomposition} (SPD) and \emph{SP labeling} (SPL), respectively. To this end, each RIS element modulates the incoming SP's phase by shifting it by one of the values in the \emph{phase shift profile} (PSP) lists satisfying resolution requirements. By interworking with a conventional \emph{orthogonal frequency division multiplexing} (OFDM) waveform, the user equipment can construct a 2D spectrum map that couples each SP’s \emph{time-of-arrival} (ToA) and PSP.
Then, we design SPL by mapping SPs with the corresponding reflected RIS elements when they share the same PSP. Given two unlabeled SPs, we derive a geometric discriminant from checking whether the current label is correct. It can be extended to more than three SPs by sorting them using pairwise geometric discriminants between adjacent ones. From simulation results, it has been demonstrated that the proposed 2D-SPC achieves consistent localization accuracy even if insufficient PSPs are given.
\end{abstract}
\section{Introduction}
\label{Sec:I}

Precise localization is one of the essential requirements for 6G communications other than a high data rate, demanding ten times higher accuracy than a 5G counterpart \cite{bourdoux20206g}.
Consideration of new frequency bands in millimeter-wave (mmWave) and terahertz ensures higher usable bandwidth and frequency, potentially leading to higher location accuracy \cite{8421288, 8421291}.
However, the high frequency gives new challenges in terms of coverage and reliability, as signals can be blocked by obstacles and may not be sufficient to ensure adequate coverage in non line-of-sight (NLoS) channel conditions. To tackle this problem, \emph{reconfigurable intelligent surface} (RIS) is expected as a
potential key technology for precise localization by providing controllable supplementary links. Due to its highly flexible characteristics in electromagnetic (EM) field management, it offers new potential for mobile user equipment (UE) localization in the presence of only one base station (BS), often called an anchor node.

In literature, various RIS-assisted localization methods have been proposed in \emph{far-field} (FF). The effect of RIS on positioning accuracy in FF has been studied by deriving the fundamental performance limits based on \emph{Cramér-Rao bound} (CRB) in multiple-input-multiple-output (MIMO) systems \cite{9129075}. When the LoS path is blocked, RIS phase shift configuration has been proposed in \cite{9807280,9500437} to develop the accuracy of estimating channel parameters for precise localization. 
The above methods assume that the UE is equipped with an antenna array, while \cite{9782100} tackles the localization problem by the \emph{time-of-arrival} (ToA) and angle of departure (AoD) in a single antenna receiver system with LoS path.
On the other hand, single-anchor localization
can also be achieved by deploying multiple RISs \cite{9215972}.
Extending to the multi-user scenario, \cite{9528041} have proposed joint localization and RIS phase design using ToA measurements in multiple RISs enabled users.
 The limitation of FF localization is the requirement of multiple antennas or multiple RISs to localize UEs.


Due to RIS's large aperture, radio signals' propagation trajectories via RIS, referred to as \emph{signal paths} (SPs) throughout this work, can change significantly depending on the location at which the signals are reflected on RIS. This phenomenon gives rise to a \emph{near-field} (NF) propagation \cite{alexandropoulos2020reconfigurable}.
Localization performance of \cite{9129075, 9807280, 9500437, 9782100, 9215972, 9528041} can be limited if the FF assumption does not hold, which can occur when the RIS's aperture size $D$ is significantly larger than the concerned wavelength $\lambda$.
To be specific, \emph{Fraunhofer distance} (FD) is the radiation boundary between the NF and FF regions, given as $2D^2/\lambda$ \cite{rappaport1996wireless,balanis2015antenna}. The distance less than the boundary is not in FF but in the NF region. For the prototype of RIS, the FD is approximately $38$~m and $87$ m at $2.3$ GHz \cite{9020088} and $5.8$ GHz \cite{9551980}, respectively, for a RIS of size $D=1.5$ m. 
This means most indoor and short-range communication systems should be designed in the NF rather than the FF regime.

Unlike the FF condition, SP's flight times reflected from the large RIS are different due to the characteristics of spherical waves. It provides new opportunities to estimate the UE's location using only a single RIS in a single-input-single-output (SISO) system.
Generally, there are two types of algorithms of RIS-assisted NF localization as below: i) direct localization and ii) two-stage localization.
First, direct localization can solve the problem based on the maximum likelihood (ML) estimator when the signal model is given mathematically, so precise localization can be performed. So, most of the RIS-assisted NF localization in the literature focuses on direct localization \cite{9860413, 10001209, 10017173,9148961,9650561}. Authors in \cite{9860413} first derived the closed form of the Fisher information matrix (FIM) and CRB of position.
In \cite{10001209}, both RIS position and orientation are estimated by a quasi-Newton method with known BS and UE. To enhance localization performance, the RIS profile can be optimized and localization is performed with optimized RIS. For example, \cite{10017173} proposes joint localization and online RIS calibration considering a realistic RIS amplitude model introduced in \cite{9148961}, which relies on equivalent circuit models of individual reﬂecting elements. For multi-target localization, a RIS phase profile is designed to concentrate on the received energy in areas of interest \cite{9650561}.

Extending to the multiple antenna scenarios, \cite{9838395} and \cite{10287134} propose a grid design to reduce the complexity of the ML estimator. Considering multipath channels, \cite{rinchi2022compressive} proposes a compressed sensing (CS) technique tailored to address the issues associated with NF localization and model mismatches. In \cite{9508872}, the CRB is derived for assessing the ultimate localization and orientation performance of synchronous and asynchronous signaling schemes. Furthermore, as in the FF counterpart, the RIS's role in the NF regime has focused on enhancing the received signal's strength favorable to the inference process. For example, RIS is configured to act as a lens receiver in \cite{9500663}, which helps remove the phase offset effect due to large RIS when a user's location with an error distribution is given in advance. Authors in \cite{9777939} explored joint NF localization and channel reconstruction with extra-large RIS-aided systems.
However, the above direct localization requires an exhaustive search to detect the ML estimator, which results in high computational complexity. Therefore, they are difficult to solve efficiently and difficult to adopt into practical usage.

While numerous RIS-assisted NF localization algorithms in the literature focus on direct localization, our main interest is two-stage localization, whose computation complexity is relatively small compared with the direct localization counterpart. When reducing our scope to two-stage localization, the SISO system is considered more challenging than the multiple-antenna systems since all reflected SPs, which are non-coherently combined at the receiver, should be decomposed without resolvability in an angular domain. One representative algorithm based on a two-stage localization is proposed in \cite{9625826}, in which a sequential phase shift at the RIS allows the single antenna receiver to decompose the composite SPs, called \emph{one-dimensional SPC} (1D-SPC).
Each SP is associated with a unique codeword, which is one-to-one mapped to the location of the RIS element that reflects the SP. This allows the ToA of each SP to be extracted without interference from other SPs. Multiple triangular geometries are then constructed from the ToAs, enabling the design of a low-complexity algorithm by leveraging classic geometric localization techniques.
    
For the practical use of SPC, the challenge due to the limited number of codewords should be addressed.
Note that the number of orthogonal codewords depends on the codeword length, leading to significant latency for the localization process if every RIS element uses an exclusive codeword. 
In other words, the number of available codewords is limited to fulfill a strict latency requirement, less than the number of RIS elements. Considering a discrete phase shift configuration of RIS elements further reduces the available number of codewords. Consequently, several SPs should be modulated by the same codeword, causing the failure of 1D-SPC due to the collisions among them.

This work aims to develop a novel yet practical SPC algorithm for NF localization that overcomes the above limitation of 1D-SPC. To this end, we jointly use phase and time domains for SPC, which is thus called \emph{two-dimensional SPC} (2D-SPC). The proposed 2D-SPC intentionally modulates the temporal domain instead of passively detecting the Doppler shift, assuming that the target is stationary. Multiple SPs can be well classified if resolved in the 2D space. For example, although several RIS elements utilize the same codeword, they can be classified in the time domain when their ToAs are significant. When the difference between a few ToAs is less than the resolution limit depending on the given bandwidth, it is possible to classify them by assigning different codewords observable in the phase domain. As a result, the proposed 2D-SPC incorporates the ToA estimation into SPC to be operable under a limited number of codewords while not degrading localization accuracy. The proposed 2D-SPC consists of two-step as follows:
\begin{itemize}
   \item \textbf{\emph{Signal Path Decomposition} (SPD) in 2D Spectrum}: As a first step of 2D-SPC, we aim to decompose all RIS-reflected SPs in the 2D spectrum map constructed as follows. Consider a conventional \emph{orthogonal frequency division multiplexing} (OFDM) waveform. Each RIS element modulates the incoming OFDM signal's phase by sequentially shifting it by one of the values in the \emph{phase shift profile} (PSP) lists. Using \emph{two-dimensional inverse discrete Fourier transform} (2D-IDFT), the received signal can be analyzed in a 2D spectrum map that couples each SP's ToA and PSP. All SPs can be decomposable if either ToAs or PSPs satisfy their resolution limits determined by the system bandwidth and the number of possible PSPs. Then, we derive the optimal PSP lists to fulfill the requirement and discuss the assignment of PSPs to RIS elements.    
    \item \textbf{\emph{Signal Path Labeling} (SPL) with Geometric Discriminant}: The second step is to label the decomposed SPs with their reflected RIS elements. The labels are automatically given when every RIS element uses an exclusive PSP without duplication. On the other hand, given an insufficient number of PSPs less than RIS elements, several RIS elements share the equivalent PSP. Since SPs using the equivalent PSP are not distinguished in the phase domain during the SPD procedure, it is impossible to know which RIS element (as an anchor) matches the corresponding estimated ToA. To address the issue, we derive the discriminant from checking whether the current label is correct by establishing the geometric relationship between two SPs in terms of their ToAs. This can be extended into SPL with more than three SPs by sorting them through pairwise discriminants between adjacent ones. In summary, we design a low-complexity SPL algorithm based on the discriminant, whose effectiveness is verified by various simulations that the localization result is consistently accurate regardless of a different number of PSPs.    
\end{itemize}

The main contributions of this work are summarized as follows:
\begin{itemize}
   \item We first design the SPD algorithm by modulating the incoming OFDM signal's phase by sequentially shifting it by one of the values in the PSP lists. Using 2D-IDFT, the received signal is analyzed in a 2D spectrum map that couples each SP's ToA and PSP. 
    \item Secondly, we propose the low-complexity SPL algorithm design with geometric discriminant given an insufficient number of PSPs. Given the current estimated position, we check whether the current label is correct by establishing the geometric relationship between two SPs in terms of their ToAs.
    \item We compare the proposed 2D-SPC against the 1D-SPC and show that the proposed designs outperform the benchmark. The effectiveness of 2D-SPC is verified by various simulations, and the localization result is consistently accurate regardless of the number of PSPs.
\end{itemize}

The rest of this paper is organized as follows. In Section~\ref{Sec:II}, we present a system model for RIS-assisted NF localization and provide an overview of the proposed 2D-SPC. Then, we define an essential design principle for perfect SPD and discuss the optimal PSP lists and their assignments in Section ~\ref{Sec:III}. Section~\ref{Sec:IV} presents the SPL using the geometric discriminant. In Section~\ref{Sec:V}, the computational complexity analysis of 2D-SPC is presented compared to 1D-SPC. Section~\ref{Sec:VI}, the effectiveness of the proposed 2D-SPC is evaluated through extensive simulations. The conclusion is finally drawn in Section \ref{Sec:VII}.

\textit{Notation:} We use boldface lower-case and upper-case characters to represent vectors and matrices, respectively. $x^{(i)}$, $\mathbf{x}^{(i)}$, $\mathbf{X}^{(i)}$, and $\mathcal{X}^{(i)}$ indicate variable, vector, matrix and set only related to the $i$-th numerology. The operators $(\cdot)^{*}$, $(\cdot)^{\top}$ and $(\cdot)^{\mathsf{H}}$ represent the conjugate, transpose and Hermitian operators, respectively. $\left[\mathbf{X}\right]_{i,j}$ and $\left[\mathbf{x}\right]_{i}$ denote $(i,j)$-th element of matrix $\mathbf{X}$ and $i$-th element of vector $\mathbf{x}$, respectively. An $N$-dimensional identity matrix is denoted as $\mathbf{I}_{N}$. $\text{diag}(\mathbf{x})$ returns a square diagonal matrix with the elements of vector $\mathbf{x}$ on its main diagonal. $\lVert\mathbf{x}\rVert$ returns a Euclidean distance of vector $\mathbf{x}$. $|\mathcal{X}|$ indicate the cardinality of set $\mathcal{X}$. $\mathbb{C}^{M}$ and $\mathbb{R}^{M}$ denote the set of $M$ complex and real vectors, respectively.

\section{System Model and Problem Definition}
\label{Sec:II}
This section considers a 3D NF localization assisted by a deployed RIS. We first describe the localization scenario and the relevant channel and signal models. Last, we provide an overview of the proposed technique to enable NF localization with RIS.

\subsection{Localization Scenario}\label{sec:Senario}
\begin{figure}[t]
\centering
\includegraphics[width=1.0\linewidth]{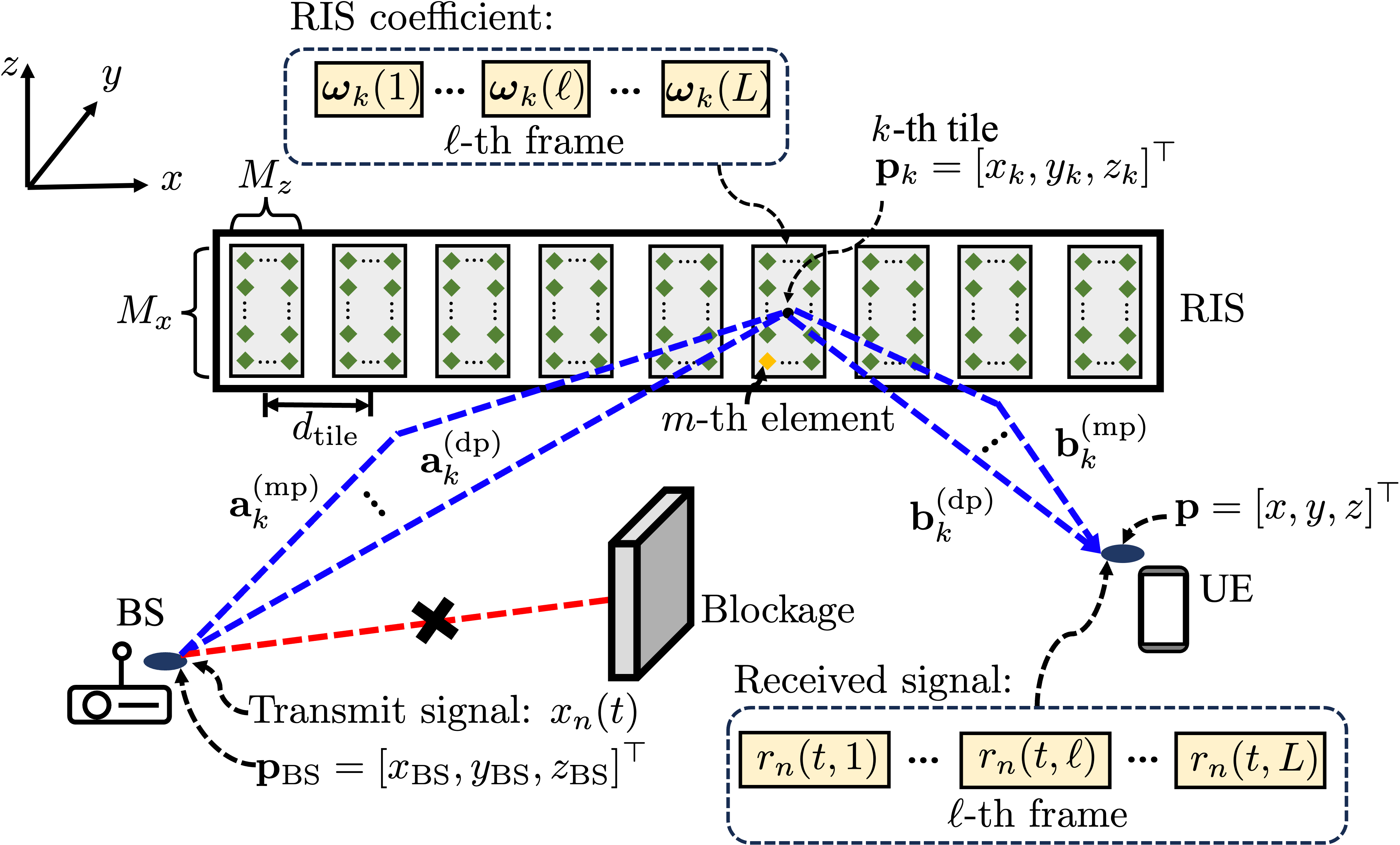}
\caption{A localization scenario comprising a pair of single antenna BS and single antenna UE and the RIS deployed in proximity. Given the locations of BS and RIS as prior information, we aim to estimate the UE's 3D location. We assume that UE is located on the ground (i.e., $z=0$).
\label{Fig: Geometric Model}}
\end{figure}

As depicted in Fig. \ref{Fig: Geometric Model}, we consider a localization scenario with a pair of single antenna BS and a single antenna UE, assisted by a RIS deployed in proximity. The RIS comprises $K$ tiles, each of which is provisioned with $M$ elements arranged in a rectangular array of $M_x \times M_z$ with half-wavelength spacing (i.e., $M=M_x \times M_z$). Adjacent RIS tiles are separated by a distance $d_{\text{tile}}$. There exists an obstacle between the BS and UE, blocking a LoS path, and only NLoS paths via RIS are available\footnote{This is not a limiting assumption but only invoked for notation convenience. When the LoS path is present, it can be separated from the RIS path as in \cite{9625826}.}. All entities' locations, assumed to be stationary, are defined in the same global reference coordinates system. Specifically, the 3D coordinates of BS and UE are denoted by $\mathbf{p}_{\text{BS}}=[x_{\text{BS}},y_{\text{BS}},z_{\text{BS}}]^\top \in \mathbb{R}^{3}$ and $\mathbf{p}=[x,y,z]^\top \in \mathbb{R}^{3}$, respectively. We assume that UE is located on the ground (i.e., $z=0$). For RIS, the coordinates of the $k$-th tile's center and the $m$-th element belonging to the $k$-th tile are $\mathbf{p}_{k}=[x_{k},y_{k},z_{k}]^\top \in \mathbb{R}^{3}$ and  $\mathbf{p}_{k,m}=[x_{k,m},y_{k,m},z_{k,m}]^\top \in \mathbb{R}^{3}$, respectively. Estimating the UE's position, say $\mathbf{p}$, is a primary goal in this work. Without loss of generality, the other locations except $\mathbf{p}$ are assumed to be given in advance.

\subsection{Channel Models}
We consider a downlink SISO channel model comprising $K$ NLoS paths reflected by RIS. The NLoS paths can be modeled as a cascade channel from the BS to the UE via a RIS tile. Assume that both BS and RIS are perfectly synchronized with each other. From the perspective of the $k$-th RIS tile, the $k$-th forward channel $\mathbf{a}_k=[a_{k,1},\cdots,a_{k,M}]^\top \in \mathbb{C}^{M \times 1}$ whose $m$-th entry represents the channel from the BS to the $m$-th element, given as
\begin{align}
a_{k,m} = a_{k,m}^{(\text{dp})}+a_{k,m}^{(\text{mp})},
\label{eq: array vector_BR_total}
\end{align}
where $a_{k,m}^{(\text{dp})}$ and $a_{k,m}^{(\text{mp})}$ representing the direct and multipath of forward components, respectively. The direct path component $a_{k,m}^{(\text{dp})}$ is given by
\begin{align}
a_{k,m}^{(\text{dp})} =\frac{\lambda}{4\pi \left\|\mathbf{p}_{\text{BS}} -\mathbf{p}_{k} \right\|} \exp \left( - j  \frac{{2\pi }}{\lambda} \left\|\mathbf{p}_{\text{BS}} -\mathbf{p}_{k,m} \right\|\right),
\label{eq: array vector_BR-DP}
\end{align}
where attenuation factor $\lambda/(4\pi \left\|\mathbf{p}_{\text{BS}} -\mathbf{p}_{k} \right\|)$ with the wavelength $\lambda$. We assume the $J$ multipath model, which is widely used in multipath simulations, given by
\begin{align}
a_{k,m}^{(\text{mp})} =\sum_{j=1}^{J}\epsilon_{j} \exp \left\{ - j  \frac{2\pi}{\lambda} (\left\|\mathbf{p}_{\text{BS}} -\mathbf{p}_{k,m} \right\|+d_{j})\right\},
\label{eq: array vector_BR-MP}
\end{align}
where $\epsilon_{j}$ and $d_{j}$ are random variables expressing, respectively, the amplitude and additional length of the $j$-th path.

Similarly, backward channel $\mathbf{b}_k=[b_{k,1},\cdots,b_{k,M}]^\top\in \mathbb{C}^{M \times 1}$ whose $m$-th entry represents the channel from the $m$-th element, given as
\begin{align}
b_{k,m} = b_{k,m}^{(\text{dp})}+b_{k,m}^{(\text{mp})},
\label{eq: array vector_RU_total}
\end{align}
with the direct component $b_{k,m}^{(\text{dp})}$ given by
\begin{align}
b_{k,m}^{(\text{dp})} =\frac{\lambda}{4\pi  \left\|\mathbf{p} -\mathbf{p}_{k} \right\|} \exp \left( - j  \frac{{2\pi }}{\lambda } \left\|\mathbf{p} -\mathbf{p}_{k,m} \right\|+j\varphi_0\right),
\label{eq: array vector_RU-DP}
\end{align}
where the attenuation factor $\lambda/(4\pi  \left\|\mathbf{p} -\mathbf{p}_{k} \right\|)$ and the phase offset $\varphi_0$. The multipath component $b_{k,m}^{(\text{mp})}$ of the backward channel is modeled according to the statistical model as in \eqref{eq: array vector_BR-MP}.

Consider multi-frame transmissions comprising $L$ frames. The configuration profile of the $k$-th RIS tile, denoted by $\boldsymbol{\omega}_{k}(\ell) \in{\mathbb C}^{M \times 1}$, is given as
\begin{equation}
    \boldsymbol{\omega}_{k}(\ell)= 
[\omega_{k,1}(\ell), \cdots, \omega_{k,M}(\ell)]^{\top},
\label{eq: RIS control vector}
\end{equation}
where the RIS coefficient of the $m$-th element at the $\ell$-th frame, denoted by ${\omega}_{k,m}(\ell) \in{\mathbb C}$, is given~as
\begin{equation}
    {\omega}_{k,m}(\ell)= \gamma_{k} e^{-j \theta_{k,m}(\ell)}, \quad \ell=1,\cdots, L,
\label{eq: RIS coefficient}
\end{equation}
which is determined by a reflection coefficient $\gamma_{k}\in \{0,1\}$ and a phase shift $\theta_{k,m}(\ell)\in[0,2\pi)$. By combining \eqref{eq: array vector_BR_total}, \eqref{eq: array vector_RU_total} and \eqref{eq: RIS coefficient}, the resultant channel gain from the BS to the UE via the RIS's $k$-th tile at the $\ell$-th frame, denoted by $g_{k}(\ell)\in\mathbb{C}$, is given as\footnote{The inter-distance between tiles is thus relatively large, allowing us to ignore the correlation between them, as followed in several works in the literature (e.g., \cite{9625826,9508872,9500663,rinchi2022compressive}).}
\begin{align}\label{eq: channel gain_NLoS}
    g_{k}(\ell)=\mathbf{b}_{k}^{\top}\text{diag}\left[\boldsymbol{\omega}_{k}(\ell)\right] \mathbf{a}_{k}, \quad k=1,\cdots, K, \quad \ell=1,\cdots, L.
\end{align}
By following the approach in \cite{9625826}, the configuration of each element in the same tile is set to be equivalent, namely, $\omega_{k,m}(\ell)=\omega_k(\ell)=\gamma_k e^{j \theta_k(\ell)}$ for all $m$. In other words, the configuration profile of \eqref{eq: RIS control vector} is reduced to $\omega_k \boldsymbol{1}_M$.


\subsection{Transmit and Receive Signal Models}
\label{sec:II-C}
Consider an OFDM signal as a localization waveform \cite{9625826}. The BS broadcasts the OFDM signal $\mathbf{x}(t)=\left[x_{1}(t), \ldots, x_{N}(t)\right]^{\top} \in \mathbb{C}^{N}$ across a set of $N$ sub-carriers. 
With frequency spacing $\delta$ and the frame duration $T=1/\delta$, 
the $n$-th sub-carrier's baseband signal $x_{n}(t)$ is given as
\begin{align}
    x_{n}(t)=\sqrt{\frac{P}{N}}e^{j 2\pi f_{n} t}, \quad 0\leq t<T,
\end{align}
where $P$ is a transmit power of BS and $f_{n}=f_{c}+(n-(N+1)/2)\delta$ is frequency of $n$-th sub-carrier with carrier frequency $f_c$.    
Next, given the BS-UE synchronization gap $t_0$, the ToA from the BS to the UE via the $k$-th RIS tile, denoted by $\tau_k$, is given as
\begin{equation}\label{delay}
\tau_k=
    \frac{\left\|\mathbf{p}_{\text{BS}} -\mathbf{p}_{k} \right\|}{c} +\frac{\left\|\mathbf{p} -\mathbf{p}_{k} \right\|}{c} + {t_0}, \quad k=1,\ldots,K.\;
\end{equation}
where $c$ is the light speed ($c=3 \times 10^8$ m/s).
For the $\ell$-th frame, the received signal of the $n$-th sub-carrier at time $t$, denoted by $r_{n}(t,\ell)\in \mathbb{C}$, is given as
\begin{align}
    r_{n}(t,\ell)=\sum_{k=1}^{K}  \underbrace{g_{k}(\ell) e^{-j2\pi f_n \tau_k}x_{n}(t)}_{y_{n,k}(t,\ell)}+{w}(t,\ell), \nonumber\\
    \quad 0\leq t<T, \quad \ell=1,\ldots, L,
    \label{eq: observation model_NLoS}
\end{align}
where ${w}(t,\ell)\in \mathbb{C}$ is the thermal noise following $\mathcal{CN}\left(0,N_0 \right)$ and ${y}_{n,k}(t,\ell)\in \mathbb{C}$ is the $n$-th sub-carrier's received signal reflected from the $k$-th RIS tile at the $\ell$-th frame when ignoring ${w}(t,\ell)$. We denote $\mathbf{Y}_{k}(t)$ with $\left[\mathbf{Y}_{k}(t)\right]_{n,\ell}=y_{n,k}(t,\ell)$ grouping all signals relevant to the $k$-th RIS tile, defined as SP $k$ throughout this work. The set containing all SPs in \eqref{eq: observation model_NLoS} is denoted by $\mathcal{K}$, given as $\mathcal{K}=\{1,2,\cdots, K\}$. The primary goal is to find the UE's position $\mathbf{p}$ after observing $\mathbf{R}(t)\in \mathbb{C}^{N\times L}$  with $\left[\mathbf{R}(t)\right]_{n,\ell}=r_{n}(t,\ell)$ of \eqref{eq: observation model_NLoS}. To achieve this goal, the received signal $\mathbf{R}(t)$ must be decomposed into $K$ SPs, and the corresponding ToAs $\{\tau_k\}$ must be extracted. Then, each observed ToA should be labeled with the RIS element that is supposed to be the ground truth.

\section{Two-Dimensional Signal Path Decomposition using Phase Modulation}
\label{Sec:III}
As the first step of SPC, this section tackles SPD in Sec. \ref{subsection3_1} by proposing a novel RIS configuration called \emph{phase modulation} (PM). Especially, PM enables the UE to construct a 2D spectrum map by transforming the signal's dimensions into time and phase domains. Then, we define an essential design principle for perfect SPD and discuss the optimal PSP lists and their assignments. Last, we extract each SP's ToA from the 2D spectrum map.

\subsection{Definition of 2D-SPC}\label{subsection3_1}
\begin{figure}[t]
\centering
\includegraphics[width=0.85\linewidth]{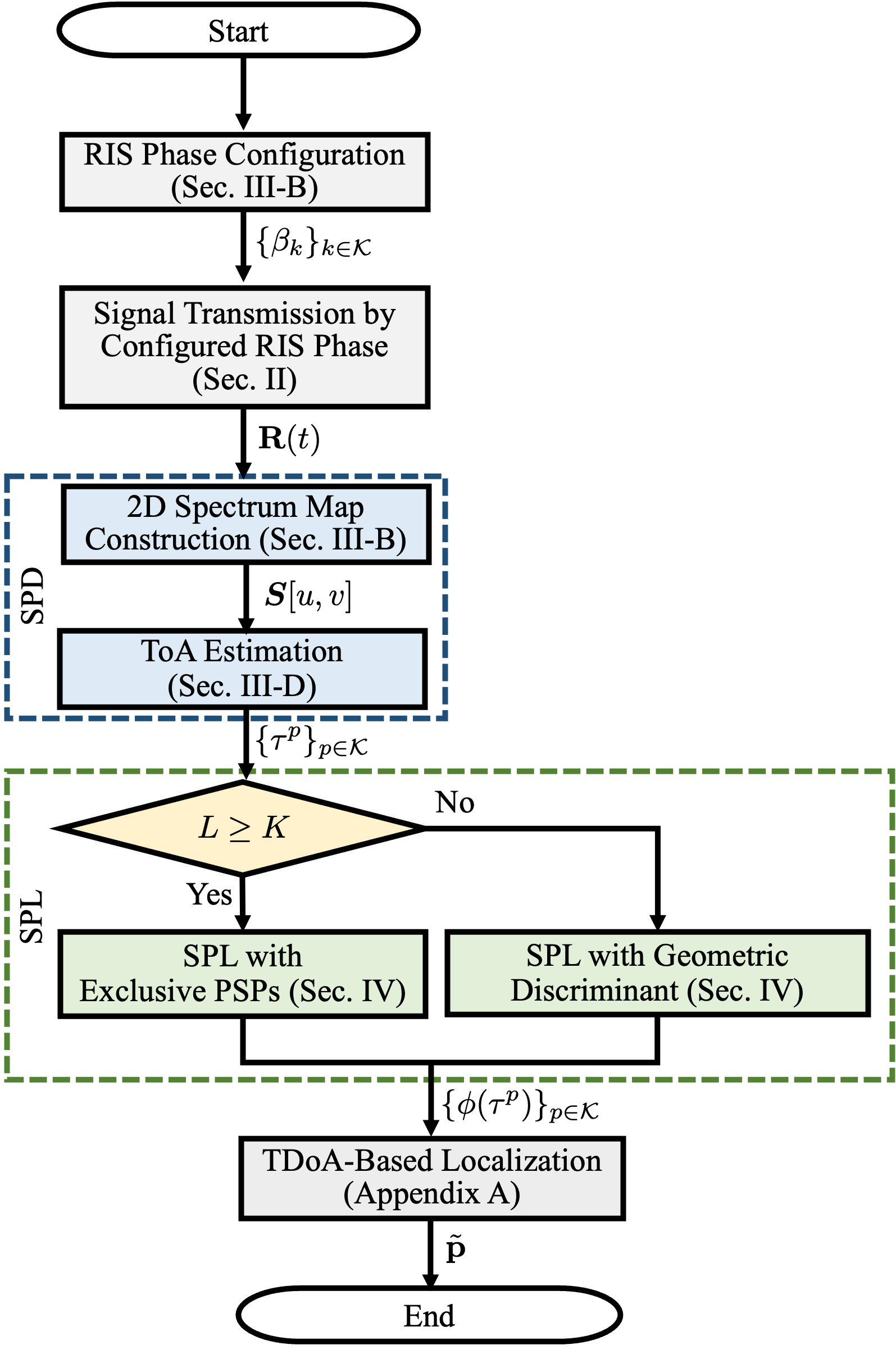}
\caption{A flow chart of the proposed 2D-SPC comprising SPD and SPL. Their input-output relations between each step are specified.}\label{Fig: Flow-Chart}
\end{figure}

Recall that SP is the radio signals' propagation trajectories via RIS, which change significantly depending on the location at which the signals are reflected on RIS. We use a two-step approach in that every SP is classified first, and the UE's location is estimated next. The second step is straightforward by applying the existing method summarized in Appendix A of the extended version. On the other hand, our core interest is in the first step, called SPC, of which a detailed definition is given below.
\begin{definition}[Signal Path Classification]\label{Def1} Given the superimposed signal $\mathbf{R}(t)$ across a set of $N$ sub-carriers and $L$ frames of \eqref{eq: observation model_NLoS}, SPC comprises the following two procedures:  
\begin{enumerate}
    \item \textbf{Signal Path Decomposition}: The objective of SPD is to decompose SPs, say $\{\mathbf{Y}_{k}(t)\}$, to extract their ToAs $\{\tau_{k}\}$. SPD is said to be perfect if all $K$ SPs are decomposable with precise ToA estimations without interference from the other SPs. The detailed process of SPD will be explained in Sec. \ref{Sec:III}. 
    \item \textbf{Signal Path Labeling}: For the decomposed SP $k$'s ToA $\tau_k$, the UE can be blind to the RIS element $k$ from which it was reflected, defined as its ground-truth label. SPL aims to label each observed ToA with the RIS element that is supposed to be the ground truth. We will use the index $p$ in the sequel to indicate the unlabeled SP's ToA differently from its ground truth $k$, i.e., $\tau^{p}$ being the ToA of the $p$-th SP given unlabeled. Denote $\phi(\tau^{p})$ and $\phi^*(\tau^{p})$ the estimated and ground truth labels, respectively. SPL is said to be perfect when $\phi(\tau^{p})=\phi^*(\tau^{p})$ for every $\tau^p$. The detailed process will be explained in Sec. \ref{Sec:IV}.   
\end{enumerate}
The overall flow of the proposed SPC is illustrated in Fig. \ref{Fig: Flow-Chart}.
\end{definition}

\subsection{2D Spectrum Map Construction with Phase Modulation}
\label{Sec:III-A}

PM is referred to as the RIS's tile-wise sequential configuration such that each RIS tile keeps shifting the incoming OFDM signal's phase over frames according to the predetermined PSP. When the $\ell$-th OFDM waveform is reflected via the $k$-th RIS tile, the corresponding phase shift, denoted by $\theta_k(\ell)$, is configured as
\begin{align}
    \theta_k(\ell)=2 \pi \beta_k \ell,\label{LPM}
\end{align}
where $\beta_k$ is the $k$-th RIS tile's PSP, which facilitates labeling resolved signals explained in the sequel\footnote{The target is assumed to be stationary, and no Doppler effect exists. Otherwise, additional techniques, such as the waveform redesign \cite{8815450}, should be incorporated to cancel out the Doppler shift due to high mobility, which remains our further study.}.
The distant RIS tiles' phases are modulated to increase linearly with different PSPs from the others. For the $\ell$-th frame, the received signal of the $n$-th sub-carrier, say $r_{n}(t,\ell)$, is then rewritten~as
\begin{align}
    r_{n}(t,\ell)=\sqrt{\frac{P}{N}}\left(\sum_{k=1}^K c_{k} e^{-j2 \pi \beta_k \ell} e^{j 2\pi f_n (t-\tau_k)}\right)+{w}(t,\ell),\label{r_n^t}
\end{align}
where $c_k=\mathbf{b}_{k}^{\top}\mathbf{a}_k$ for $k=1,\ldots,K$.
To cancel out the term relevant to time $t$, the UE demodulates the received signal 
by multiplying the conjugate of \eqref{r_n^t} with the corresponding transmit waveform $x_{n}(t)=\sqrt{P/N}e^{j 2\pi f_n t}$. The demodulated signal, denoted by $s_{n}(\ell)$, is given~as
\begin{align}
    s_{n}(\ell)&=\frac{1}{T}\int_{0}^{T} r_{n}(t,\ell)^{*} x_{n}(t) dt\nonumber\\&=\frac{P}{N}\left(\sum_{k=1}^{K} c_k^{*} e^{j 2\pi f_n \tau_k}e^{j  2\pi\beta_k \ell}\right) + \bar{w}(\ell),
\end{align}
\begin{figure*}[t]
\centering
\begin{subfigure}{1.0\columnwidth}
\centering
       \includegraphics[width=0.8 \linewidth]{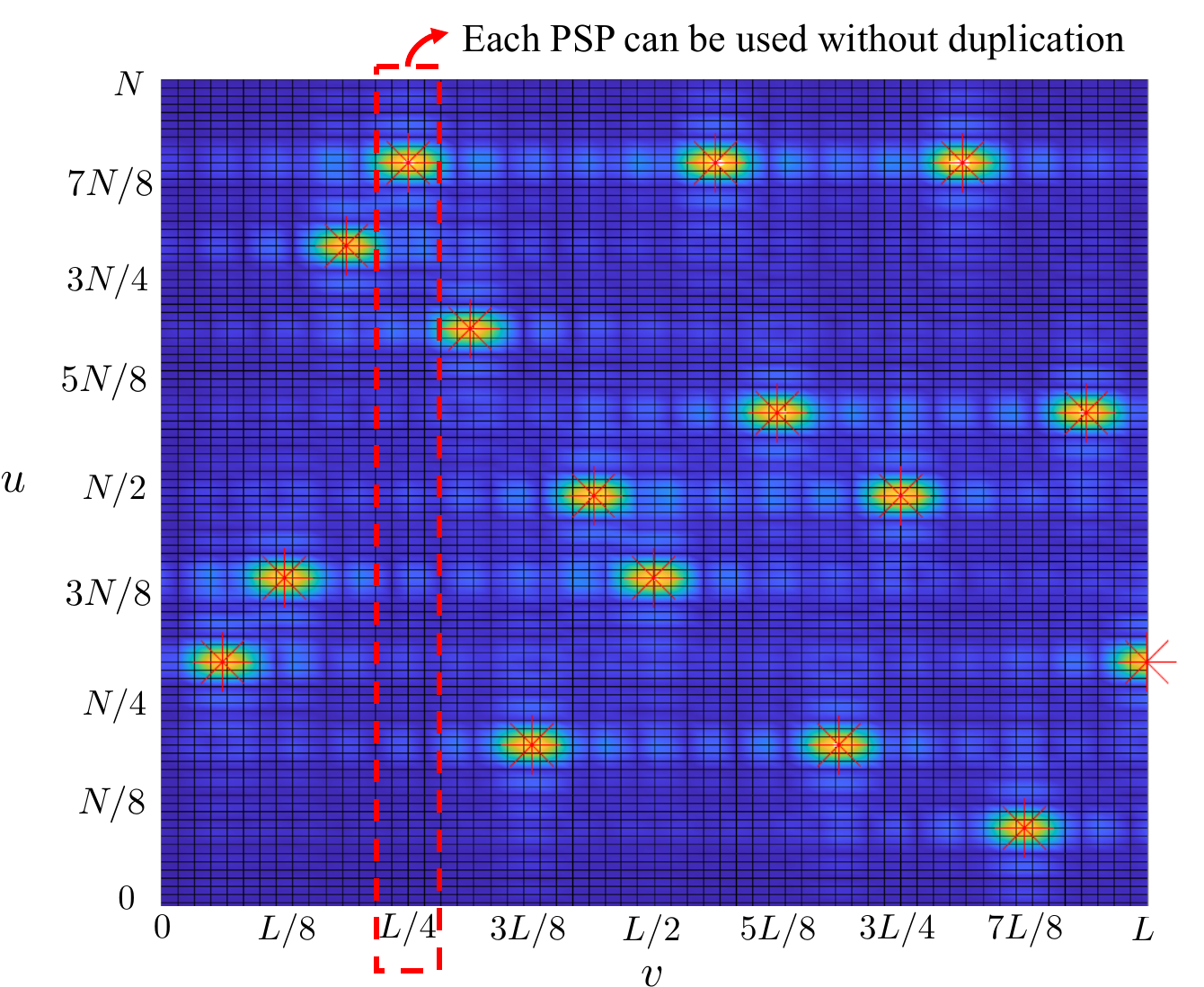}
       \caption{Sufficient PSPs ($L\geq K$)}
       \label{fig:2D-FT1}
\end{subfigure}\hfill%
\begin{subfigure}{1.0\columnwidth}
\centering
       \includegraphics[width=0.8\linewidth]{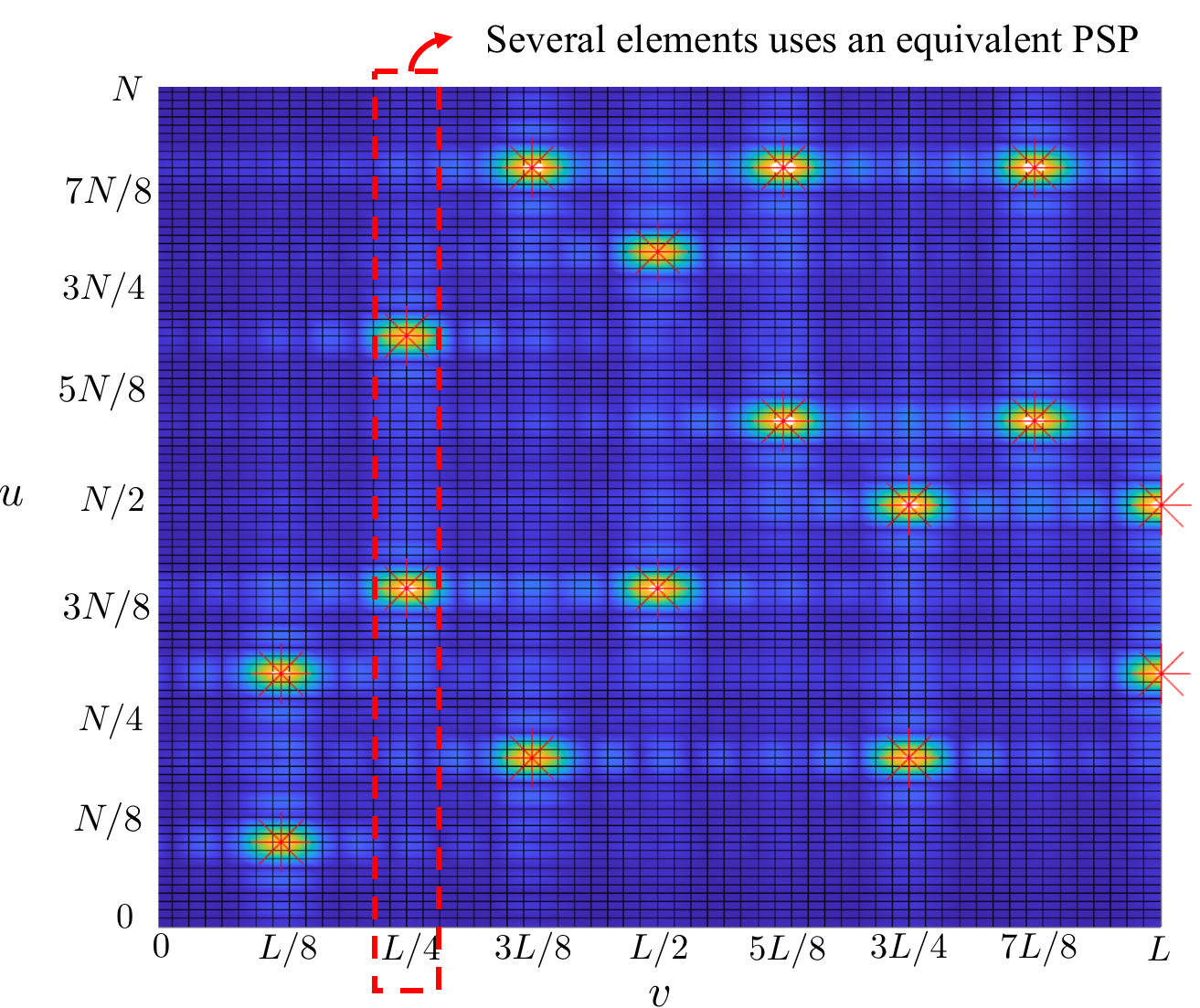}
       \caption{Insufficient PSPs ($L<K$)}
       \label{fig:2D-FT2}
\end{subfigure}
\caption{Graphical examples of the proposed 2D spectrum map under (a) sufficient PSPs ($K=L=16$) and (b) insufficient PSPs ($K=16$, $L=8$).
}
\label{Fig: 2D-IDFT}
\end{figure*}
where the noise $\bar{w}(\ell)\sim \mathcal{CN}(0,PN_0/N)$.
Since the term inside the summation is composed of two exponential functions, we can translate it into the domains of $\tau$ and $\beta$ using 2D-IDFT, given~as
   \begin{align}
    &\mathcal{S}(\tau,\beta)= \sum_{\ell=1}^{L}\bigg\{ \sum_{n=1}^{N}\left(s_{n}(\ell) e^{-j 2\pi \frac{n}{N}\tau}\right)\bigg\} e^{-j 2\pi \frac{\ell}{L}\beta} \nonumber\\
    &=\frac{P}{N}\left\{e^{-j  \pi \beta L}\sum_{k=1}^{K} c_k^{*} \mathsf{sinc}(B(\tau-\tau_k))\mathsf{sinc}( L(\beta-\beta_k))\right\}+\bar{w},\label{2DFFT}
\end{align}
where $\mathsf{sinc}(x)={\sin(\pi x)}/({\pi x})$, $B$ is the system bandwidth, and $\bar{w}$ is the noise following $\mathcal{N}(0,PN_0)$.

The above result is shown to be the superposition of multiple 2D $\mathsf{sinc}$ functions, whose main lobes are located in $\{(\tau_k, \beta_k)\}$. With the aid of PM, several overlapped 1D $\mathsf{sinc}$ functions whose ToAs are within the resolution limit can be shifted on the expanded dimension of $\beta$, helping resolve them while preserving ToA information embedded therein. We specify the above explanation in the following theorem, which is the key design principle throughout the work.

\begin{theorem} [Two-Dimensional Signal Path Decomposition]\label{Theorem1} Consider the RIS with $K$ tiles, whose phase shift configurations follow \eqref{LPM}. Under a high SNR regime, every SP can be perfectly decomposable when any SP $k$ in $\mathcal{K}$ satisfies at least one condition of two explained as follows:
\begin{itemize}
    \item Given the system bandwidth $B$, the ToA gaps to the other SPs should be no less than ${1}/{B}$, namely, $ |\tau_{k}-\tau_{k_{1}}|\geq{1}/{B}$, $\forall k_{1}\in \mathcal{K}\setminus\{k\}$.   
    \item Given the number of frames $L$, the PSP gaps to the other SPs should be no less than ${1}/{L}$, namely,  $|\beta_{k}-\beta_{k_{2}}|\geq {1}/{L}$, $\forall k_{2}\in \mathcal{K}\setminus\{k\}$. 
\end{itemize}
\end{theorem}
\noindent\begin{IEEEproof}
See Appendix B in the extended version.
\end{IEEEproof}
Note that the linear phase modulation in \eqref{LPM} is optimal from the perspective of maximizing the minimum difference between PSPs, which is proved in Appendix B of the extended version.

\subsection{Design and Assignment of Phase Shift Profile}
\label{Sec:III-B}



Recall that Theorem \ref{Theorem1}'s condition is twofold. Contrary to the first condition determined by the given BS-RIS-UE geometry, the effectiveness of the second condition depends on the lists of possible PSPs and their assignments to each RIS element, the main tasks of this subsection.  

First, we attempt to find the optimal PSP lists to fulfill the second requirement of Theorem \ref{Theorem1}. Due to the periodicity of phase, all PSPs $\{\beta_k\}$ in \eqref{LPM} should be within the range of $[0, 1)$. Besides, the resolution limit in the domain $\beta$, say ${1}/{L}$, is always achievable by setting adjacent PSPs' inter-distances to be equivalent. As a result, given the number of OFDM frame $L$, which is the same as the number of possible PSPs, the possible PSPs $\alpha(i)$ are given as
\begin{align}\label{Optimal_PM_Rates}
    \alpha(i)=\frac{i}{L}, \quad i=1,2,\ldots,L.
\end{align}
Note that an SP with the PSP $\alpha(L)$ is considered \emph{unmodulated} since the resultant phase profile $\omega_k(\ell)$ of \eqref{eq: RIS control vector} always becomes $1$ for all $\ell$.
The resulting list after the exclusion thus includes $L$ PSPs, namely,
\begin{align}
\mathcal{A}=\{\alpha(i)\}_{i=1}^{L}.
\end{align}
Next, we aim to assign PSPs in $\mathcal{A}$ to each RIS element. We consider two possible cases as follows depending on the number of elements $K$.
\subsubsection{Sufficient PSPs} Consider a case when $L\geq K$. Each RIS tile can use an exclusive PSP, as shown in Fig. \ref{fig:2D-FT1}. According to the first condition of Theorem \ref{Theorem1}, it is always possible to resolve all SPs regardless of their ToAs. As a result, perfect SPD can be achievable if the following condition holds:
\begin{align}
\beta_{k_{1}}\neq \beta_{k_{2}}, \quad \forall k_{1}, k_{2} \in \mathcal{K}.
\end{align}
In other words, all SPs are decomposable and explicitly labeled by the corresponding PSPs.

\subsubsection{Insufficient PSPs}
Consider a case when $L<K$. As shown in Fig. \ref{fig:2D-FT2}, several RIS tiles should use the same PSP, causing the failure of SPD if the gap between their ToAs is within the resolution limit $1/B$. To avoid such cases as much as possible, we set each element's phase configuration $\{\beta_k\}_{k\in\mathcal{K}}$ as follows. 
\begin{enumerate}
    \item We uniformly select $K_0$ RIS elements and allow them to use exclusive PSPs without duplication. The set including these indices is denoted by $\mathcal{K}_0$. The last $K_0$ PSPs $\left\{\alpha(i)\right\}_{i=L-K_0+1}^{L}$ are assigned to the RIS elements in $\mathcal{K}_0$. This process is essential for the subsequent SPL design introduced in the sequel.  
    \item After the above process, the PSPs $\left\{\alpha(i)\right\}_{i=1}^{L-K_0}$ should be assigned to the remaining RIS tiles in $\mathcal{K}\setminus \mathcal{K}_0$, according to the following two criteria. The first is to minimize the number of RIS elements using the same PSP, while the second is that the location of RIS elements using the same PSP should be as far apart as possible. To this end, we make the circular symmetric assignment policy as follows:
    \begin{align}
    \beta_{k}=\begin{cases}
    \alpha({i}+1), & \text{if $\beta_{\kappa}=\alpha(i)$, $i\leq L-K_0$,}\\
    \alpha(1), & \text{if $\beta_{\kappa}=\alpha(L-K_0+1)$,}\\
    \end{cases}
    \end{align}
    where $\kappa\in \mathcal{K}\setminus\mathcal{K}_0$ represents the index of the RIS tile just before $k$.
\end{enumerate}
Consequently, we group RIS tiles according to the corresponding PSPs. The subset $\mathcal{L}{(i)}$ includes the indices of the RIS tile using the PSP $\alpha(i)$, namely,
\begin{align}
\mathcal{L}{(i)}= \{k\in\mathcal{K}\setminus\mathcal{K}_0| \beta_k=\alpha(i)\},
\label{eq: label set}
\end{align}
which is also considered as the set of candidate labels in the following section.


\begin{figure}[t]
\centering
\includegraphics[width=0.9\linewidth]{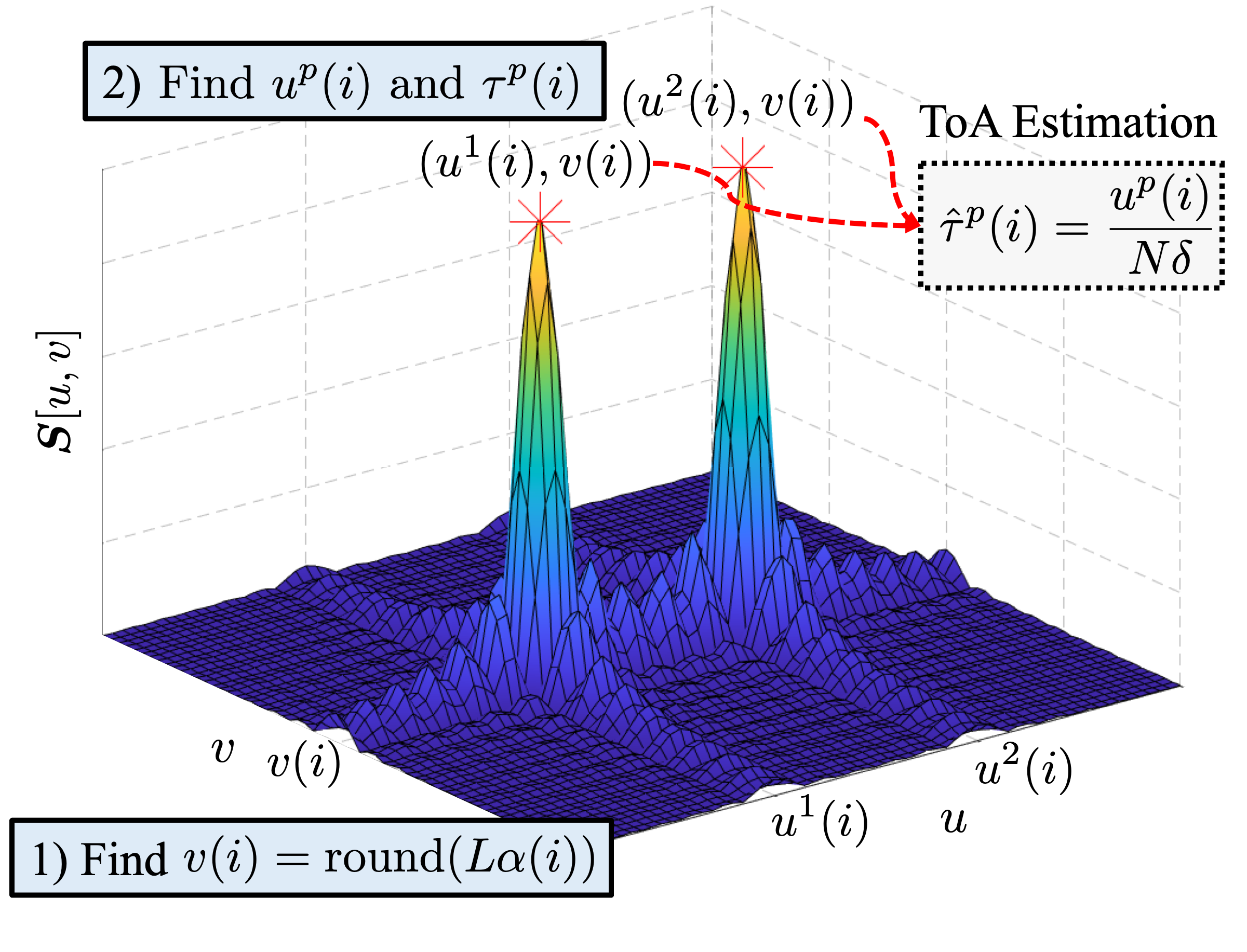}
\caption{A graphical example of ToA estimation $\{\tau^{p}(i)\}$ from $\boldsymbol{S}[u,v]$ of \eqref{eq: general 2D-IDFT} when $|\mathcal{L}(i)|=2$.\label{Fig: ToA estimation}}
\end{figure}

\subsection{ToA Estimation}
This subsection explains the detailed algorithm to estimate each decomposed SP's ToA. We introduce new variables $u$ and $v$, which are counterparts of $\tau$ and $\beta$ in the 2D-IDFT. Denote $\boldsymbol{S}\in \mathbb{C}^{N\times L}$ the result of 2D-IDFT, whose $(u,v)$-th element is given as      
\begin{align}
\boldsymbol{S}[u,v]&=\mathsf{IDFT}_{\text{2D}}\left(\{s_n(\ell)\}_{n,\ell}\right)\nonumber\\
=&\sum_{\ell=1}^L \sum_{n=1}^{N} s_n(\ell)e^{-j  \frac{2\pi n}{N}u}e^{-j  \frac{2\pi  \ell}{L}v}\nonumber\\
=&\frac{P}{N}\sum_{k=1}^{K}\sum_{n=1}^{N} \sum_{\ell=1}^{L} c_{k}^{*} e^{j 2\pi \left(\delta \tau_k-\frac{u}{N}\right)n} e^{j 2\pi  \left(\beta_k -\frac{v}{L}\right) \ell}+\bar{w}.
\label{eq: general 2D-IDFT}
\end{align}
Note that the oversampling can be adopted by replacing $N$ with $\bar{N}=Q_c N$ and oversampling factor $Q_c$ in \eqref{eq: general 2D-IDFT} and \eqref{eq: ToA set} to improve ToA resolution in IDFT, which is generally used in the literature (e.g., \cite{4533823,5535158,7820084}).
\begin{remark}
       (\emph{Effectiveness in Narrowband}) Though the proposed 2D-SPC algorithm is operable regardless of bandwidth, its effectiveness is more highlighted in narrowband scenarios. Specifically, multiple SPs can be easily separable in a two-dimensional signal space without overlapping. In general, a narrowband signal suffers from estimating precise ToAs due to its low resolution in decomposing multi-path signals. On the other hand, 2D-SPC makes it possible to estimate all SPs' ToAs without inter-path interference, reaching high accuracy beyond the Nyquist rate through an oversampling technique. 
\end{remark}

The magnitude of the 2D-IDFT, say $|\boldsymbol{S}[u,v]|$, has $K$ SP's local optimal points at $\{(u_k^*, v_k^*)\}$ when the terms inside the two exponential functions are zeros, i.e., $\delta\tau_k={u_k^*}/{N}$ and $\beta_k = {v_k^*}/{L}$. 
From the UE perspective, the detected optimal point's index, say $k$, is not mandatory when the same PSP modulates multiple SPs. Thus, a new index $p$ instead of $k$ is required to represent the detected SP. For example, as illustrated in Fig. \ref{Fig: ToA estimation}, consider SPs modulated by the PSP $\alpha(i)$, which corresponds to the local peak at $\{u^{p}(i), v(i)\}_{p\in \mathcal{L}(i)}$ and the corresponding ToAs $\{\tau^{p}(i)\}_{p\in \mathcal{L}(i)}$ can be calculated as follows:  

\begin{enumerate}
\item Find $v(i)$: Given the PSP $\alpha(i)$, the UE finds $v(i)$ as
\begin{align}
    v(i)=\mathsf{round}(L \alpha(i)), \quad i=1,2,\ldots, L,
\end{align}
where $L$ is the number of frames and $\mathsf{round}(x)$ rounds $x$ to the nearest integer.

\item Find $u^{p}(i)$ and $\tau^{p}(i)$: The set of paired indices $\{u^{p}(i)\}_{p=1}^{|\mathcal{L}(i)|}$ can be obtained by finding $|\mathcal{L}(i)|$ peaks from $\boldsymbol{S}[u,v(i)]$, which can be translated into $\hat{\tau}^{p}(i)$ as follows:
\begin{align}
        \hat{\tau}^{p}(i)=\frac{u^{p}(i)}{{N}\delta},
    \label{eq: time delay}
\end{align}
where $N$ and $\delta$ are the number of sub-carriers and their spacing specified in Sec. \ref{sec:II-C}.
\end{enumerate}
As a result, we can make the set of ToAs modulated by the PSP $\alpha(i)$, namely, 
\begin{align}
\mathcal{T}(i)=\{\hat{\tau}^{p}(i)\}_{p=1}^{|\mathcal{L}(i)|}, \quad i=1,\ldots, L,
\label{eq: ToA set}
\end{align}
where the ToAs therein are sorted in descending order, i.e., $\tau^{p}\geq \tau^{q}$ when $p<q$. Fig. \ref{Fig: ToA estimation} depicts the ToA estimation in the 2D spectrum map in the case with two SPs modulated by the same PSP.

\section{Signal Path Labeling}\label{Sec:IV}
This section aims to design SPL by finding each SP's label when they are modulated by the same PSP. To this end, we first derive a geometric discriminant from checking whether the current label is correct by establishing the geometric relationship between
two SPs regarding their ToAs. Then, we extend it into the case with more than three ToAs by sorting them through pairwise discriminants between adjacent ones.

\subsection{Geometric Discriminant}

Consider a pair of sets $\mathcal{L}$ of \eqref{eq: label set} and $\mathcal{T}$ of \eqref{eq: ToA set} associated with a typical PSP with omitting the index $i$ for simplicity. We define a \emph{degree-of-duplication} (DoD) as the number of RIS elements utilizing equivalent PSP. We start the case of DoD being two, specified by $\mathcal{L}=\{k_{1}, k_{2}\}$ and $\mathcal{T}=\{\tau^{p}, \tau^{q}\}$, where ToAs $\tau^{p}$ and $\tau^{q}$ are given as
\begin{align}
\tau^{p}=\frac{\left\|\mathbf{p}_{\text{BS}} -\mathbf{p}_{\phi(\tau^{p})} \right\|}{c}+\frac{\left\|\mathbf{p} -\mathbf{p}_{\phi(\tau^{p})}\right\|}{c}+t_0, \label{eq: ToA_k1}\\
\tau^{q}=\frac{\left\|\mathbf{p}_{\text{BS}} -\mathbf{p}_{\phi(\tau^{q})} \right\|}{c}+\frac{\left\|\mathbf{p} -\mathbf{p}_{\phi(\tau^{q})}\right\|}{c}+t_0,\label{eq: ToA_k2}
\end{align}
where $t_0$ is the synchronization gap between BS and UE, as specified in \eqref{delay}. The ToAs $p$ and $q$'s labels, say $\phi(\tau^{p})$ and $\phi(\tau^{q})$ specified in Sec. \ref{subsection3_1}, should be one-to-one mapped to the elements in $\mathcal{B}$. The ToAs of \eqref{eq: ToA_k1} and \eqref{eq: ToA_k2} are assumed to be precisely estimated by \eqref{eq: time delay} from 2D-IDFT and resolvable by satisfying the first condition in Theorem \ref{Theorem1}.  For ease of explanation, we assume $p<q$, thereby $\tau^{p}\geq \tau^{q}$ as mentioned above.

\begin{figure}[t]
\centering
\includegraphics[width=1.0\linewidth]{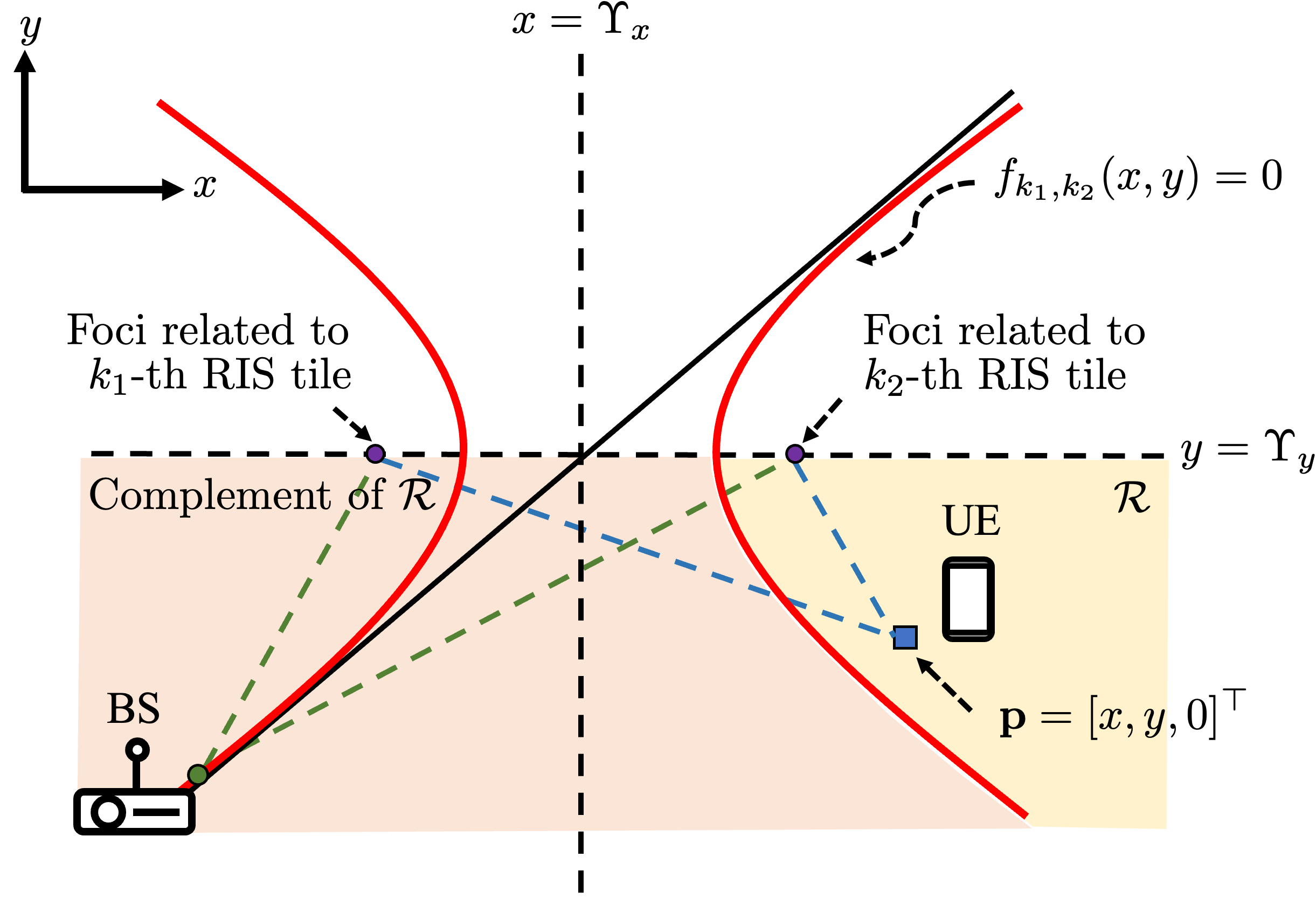}
\caption{Geometric relation between the $k_{1}$-th and $k_{2}$-th RIS tiles with the hyperbolic function of \eqref{eq: hyperbolic function}, where the $k_{1}$-th tile is assumed to be located to the left of the $k_{2}$-the tile, i.,e., $x_{k_1}\leq x_{k_2 }$. \label{Fig: Hyperbolic Function}}
\end{figure}

Let us hypothesize that $\phi(\tau^{p})=k_1$ and $\phi(\tau^{q})=k_2$. 
We aim to make the discriminant to check whether the hypothesis is correct. To this end, the relation between the ToAs $\tau_p$ and $\tau_q$ can be represented in mathematical form as follows. As shown in Fig. \ref{Fig: Hyperbolic Function}, define two lines $y=\Upsilon_y$ and $x=\Upsilon_x$ parallel and perpendicular to the RIS tile's alignment, which meet at the middle of the $k_1$-th and $k_2$-th RIS elements. The BS is assumed to be located to the left of the RIS. For notation brevity, we consider that the $k_1$-th RIS tile is always to the left of $k_2$-th RIS tile, i.e., $x_{k_{1}}<x_{k_{2}}$. By computing the difference between \eqref{eq: ToA_k1} and \eqref{eq: ToA_k2}, the region of the UE's possible positions satisfying $\tau^{p}\geq \tau^{q}$, say $\mathcal{R}$, can be defined~as
\begin{align}
    \mathcal{R}&=\left\{\mathbf{p}\in \mathbb{R}^2\mid \tau^{p}-\tau^{q}\geq 0, \phi(\tau^{p})=k_1, \phi(\tau^{q})=k_2 \right\}\nonumber\\&
    =\left\{\mathbf{p}\in \mathbb{R}^2\mid \tau_{k_1}-\tau_{k_2}\geq 0\right\}\nonumber\\&
    =\left\{\mathbf{p}\in \mathbb{R}^2\mid d_{k_1}-d_{k_2} \geq  d_{{k_2}}^{\text{BS}}-d_{{k_1}}^{\text{BS}}\right\},
    \label{eq: set of hyperbolic}
\end{align}
where $d_{k_i}={\left\|\mathbf{p} -\mathbf{p}_{{k_{i}}} \right\|}/{c}$ and $d_{k_i}^{\text{BS}}={\left\|\mathbf{p}_{\text{BS}} -\mathbf{p}_{{k_{i}}}\right\|}/{c}$ for $i\in \{1,2\}$, respectively. The boundary of $\mathcal{R}$, equivalent to the equality condition of \eqref{eq: set of hyperbolic}, can be represented as a hyperbolic curve, summarized in the following theorem.

\begin{theorem} [Geometric Discriminant]\label{Theorem2} 
Consider two unlabeled SPs' ToAs $\tau^{p}$ and $\tau^{q}$ where $\tau^{p}>\tau^{q}$. Given $\mathcal{L}=\{k_{1}, k_{2}\}$, define a hyperboloid of two sheets $f_{k_{1},k_{2}}(x,y,z)=0$ with the function $f_{k_{1},k_{2}}(x,y,z)$ defined~as \cite{reynolds1993hyperbolic}
\begin{equation}
f_{k_{1},k_{2}}(x,y,z)=\frac{(x-\Upsilon_x)^2}{\upsilon_x^2}-\frac{(y-\Upsilon_y)^2}{\upsilon_y^2}-\frac{(z-\Upsilon_z)^2}{\upsilon_z^2}-1,
    \label{eq: hyperboloid of two sheet function}
\end{equation}
where
\begin{align}
    &\Upsilon_x=(x_{k_{1}}+x_{k_{2}})/2,\nonumber\\
    &\Upsilon_y=(y_{k_{1}}+y_{k_{2}})/2,\nonumber\\
    &\Upsilon_z=(z_{k_{1}}+z_{k_{2}})/2,\nonumber\\
    &\upsilon_x=\bigg\{\sqrt{(x_{k_1}-x_{\text{BS}})^2+(y_{k_1}-y_{\text{BS}})^2+(z_{k_1}-z_{\text{BS}})^2}\nonumber\\ \quad &-\sqrt{(x_{k_2}-x_{\text{BS}})^2+(y_{k_2}-y_{\text{BS}})^2+(z_{k_2}-z_{\text{BS}})^2}\bigg\}/2,\nonumber\\
    &\upsilon_y=\upsilon_z=\\&\sqrt{\left\{(x_{k_1}-x_{k_2})^2+(y_{k_1}-y_{k_2})^2+(z_{k_1}-z_{k_2})^2\right\}/4-\upsilon_{x}^2}.\nonumber
\end{align}
We assumed that the UE is located on the ground ($z=0$). By projecting \eqref{eq: hyperboloid of two sheet function} on the xy plane, the hyperbolic function $f_{k_{1},k_{2}}(x,y)$ can be defined as
\begin{equation}
f_{k_{1},k_{2}}(x,y)=\frac{(x-\Upsilon_x)^2}{\upsilon_x^2}-\frac{(y-\Upsilon_y)^2}{\upsilon_y^2}-\frac{\Upsilon_z^2}{\upsilon_z^2}-1.
    \label{eq: hyperbolic function}
\end{equation}
Then, as illustrated in Fig. \ref{Fig: Hyperbolic Function}, the region $\mathcal{R}$ is equivalent to the area determined by the right side hyperbolic curve, given as
\begin{align}
\mathcal{R}&= \{\mathbf{p}\in \mathbb{R}^2| f_{k_{1},k_{2}}(x,y)\ge 0, x\ge \Upsilon_x, y\leq \Upsilon_y\}.
\label{eq: Region}
\end{align}
As a result, the UE can label the SPs $p$ and $q$ as $k_1$ and $k_2$, respectively, if the UE is in $\mathcal{R}$, and vice versa otherwise. 
\end{theorem}
\begin{IEEEproof}
See Appendix C in the extended version.
\end{IEEEproof}

\begin{figure*}%
\centering
\begin{subfigure}{1.0\columnwidth}
\centering
        \includegraphics[width=0.93\linewidth]{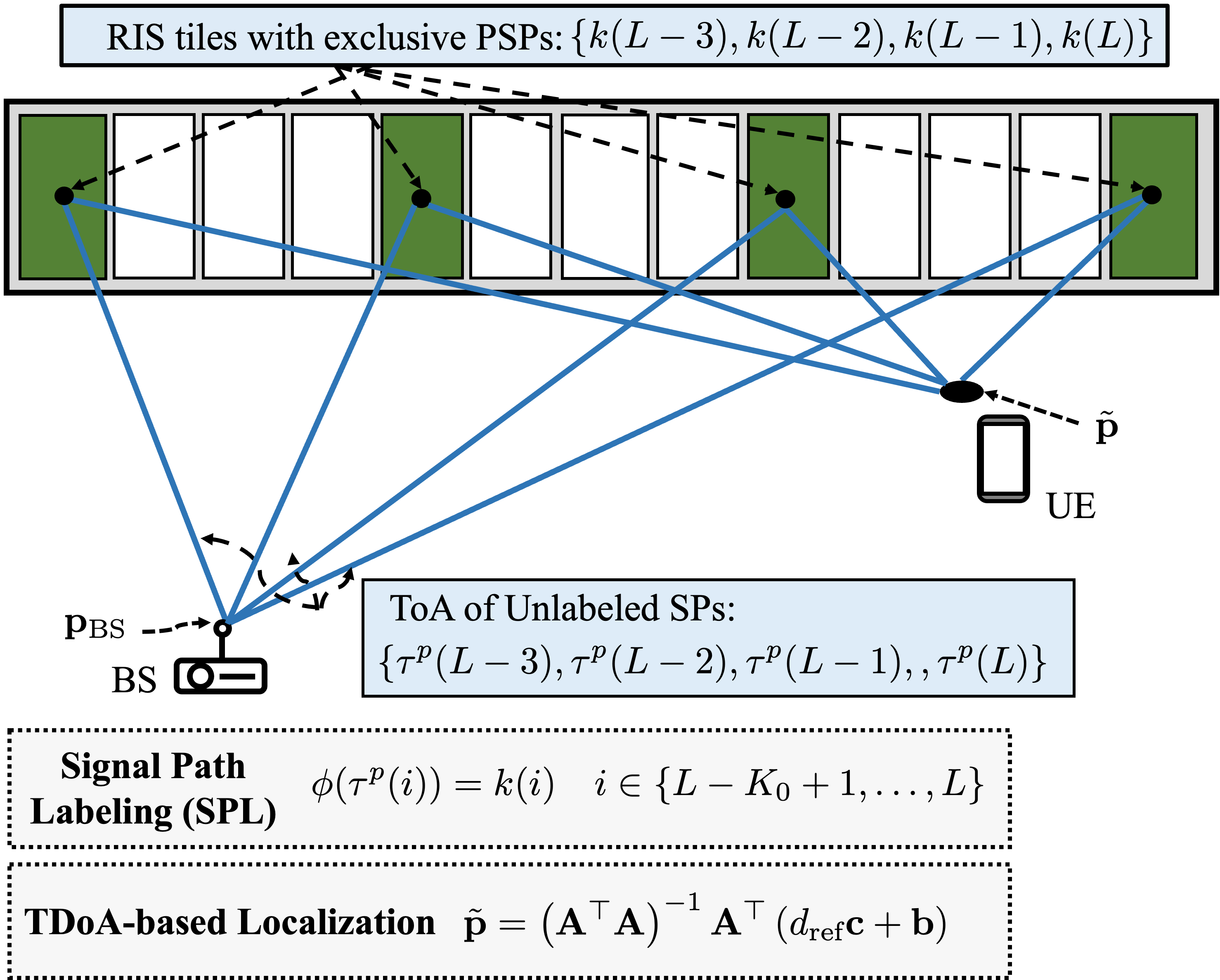}
        \caption{SPL with Exclusive PSPs (DoD is one)}
        \label{Fig: Initialization}
\end{subfigure}\hfill
\begin{subfigure}{1.0\columnwidth}
\centering
        \includegraphics[width=1.0\linewidth]{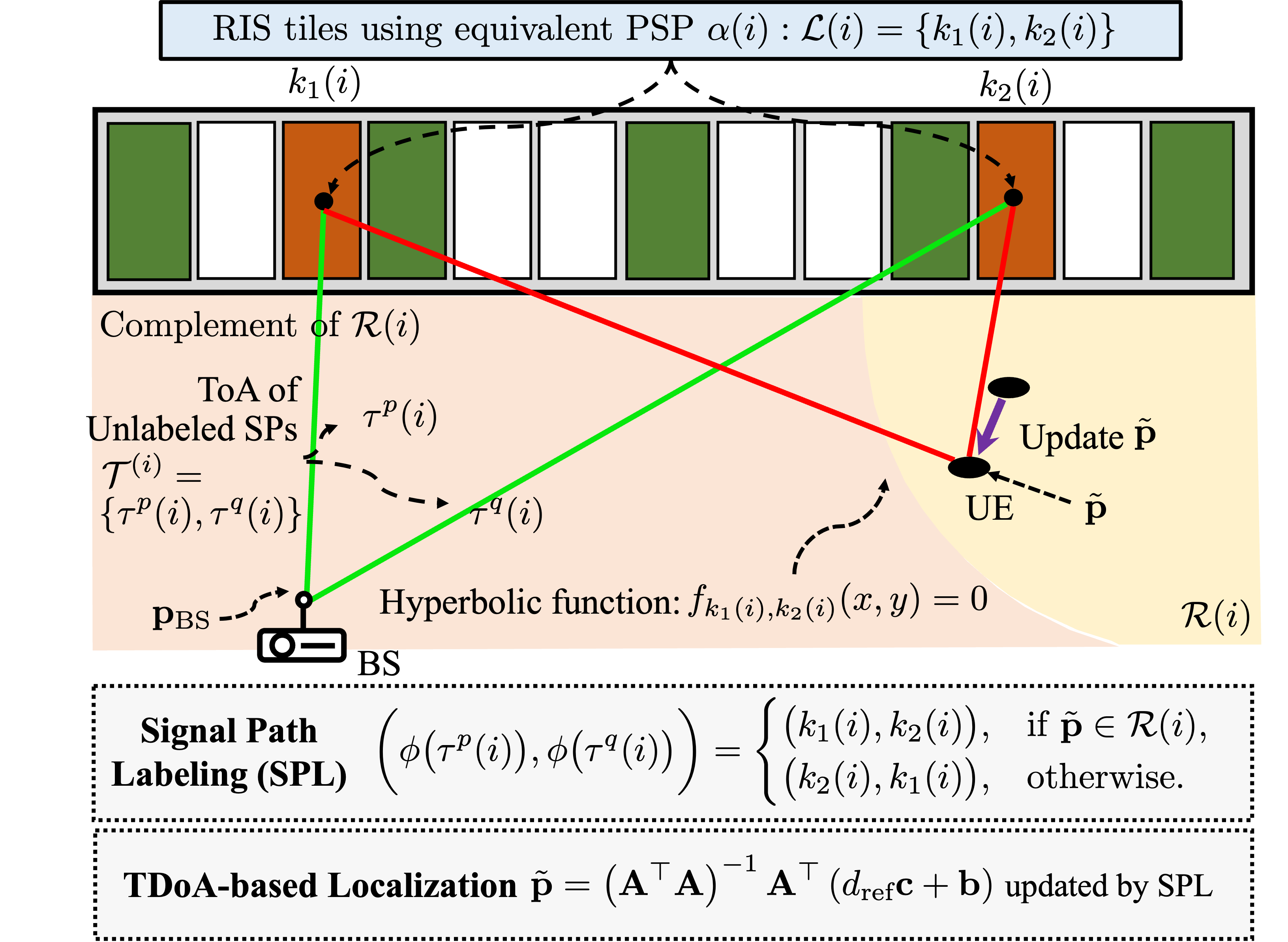}
        \caption{SPL with Geometric Discriminant (DoD is two)}
        \label{Fig: Greedy}
\end{subfigure}%
\caption{The graphical example of the proposed SPL algorithm when ${K}_{0}=4$.}
\label{fig: procedure}
\end{figure*}
It is worth noting that the discriminant in Theorem \ref{Theorem2} requires the UE's location $\mathbf{p}$, which is yet infeasible. Instead, we design an SPL algorithm in a greedy manner such that the previously estimated location of the UE, denoted by $\tilde{\mathbf{p}}$, is used as an input of the discriminant, namely, 
\begin{equation}
    \left(\phi(\tau^{p}),\phi(\tau^{q})\right)=
    \begin{cases}
    \left(k_1,k_2\right), & \text{if } \tilde{\mathbf{p}}\in \mathcal{R},\\
    \left(k_2,k_1\right), & \text{otherwise,}
    \end{cases}
    \label{eq:labeling}
\end{equation}
where the parentheses represent the ordered sequences throughout the paper\footnote{When the BS's location is on the right side of the RIS, two conditions in the geometric discriminant of \eqref{eq:labeling} need to be switched, while the others are equivalent.}.
The detailed algorithm will be elaborated in the following subsection. 

\subsection{Algorithm Description}
\label{sec:IV-C}
This subsection introduces the proposed SPL designed based on the geometric discriminant in Theorem \ref{Theorem2}. We group RIS elements as $\{\mathcal{L}{(1)}, \mathcal{L}{(2)}, \cdots, \mathcal{L}{(L)}\}$, where the subset $\mathcal{L}{(i)}$ specified in \eqref{eq: label set}, can be unfolded as
\begin{align}
\mathcal{L}{(i)}=\left\{k_1{(i)}, k_2{(i)}, \cdots, k_{|\mathcal{L}^{(i)}|}{(i)}\right\}.\label{eq: L}
\end{align}
Similarly, the sequence of unlabeled SPs' ToAs is expressed as $\left\{\mathcal{T}{(1)}, \mathcal{T}{(2)}, \cdots, \mathcal{T}{(L)}\right\}$, where the subset $\mathcal{T}{(i)}$ specified in \eqref{eq: ToA set} is unfolded as
\begin{align}
\mathcal{T}{(i)}=\left\{\tau^{p_{1}}(i),\tau^{p_{2}}{(i)},\ldots, \tau^{p_{|\mathcal{L}{(i)}|}}{(i)}\right\}. \label{eq: T}
\end{align}

The entire SPL problem can be divided into multiple sub-problems such that the labels in $\mathcal{L}{(i)}$ are one-to-one mapped to the ToAs in $\mathcal{T}{(i)}$. We sequentially solve these sub-problems in the ascending order of DoD. The detailed algorithm is summarized in Algorithm \ref{Geometric Signal Path Labeling} and explained as follows.

\subsubsection{SPL with Exclusive PSP (DoD is one)}
Consider a PSP whose DoD is~$1$, specified as
\begin{align}
\mathcal{L}{(i)}=\{k{(i)}\}, &\quad \mathcal{T}{(i)}=\{\tau^{p}{(i)}\},\nonumber\\ &\quad i\in \{L-K_0+1, \ldots, L\} 
\end{align}
Note that $i\in \{L-K_0+1, \ldots, L\} $ is allocated for PSPs whose DoD is one, as specified in Sec. \ref{Sec:III-B}. Since only one element in the above set exists, the label is explicitly given without ambiguity, namely, 
\begin{align}
\phi(\tau^{p}{(i)})=k{(i)}.
\label{eq:initial labeling}
\end{align}
As stated in Sec. \ref{Sec:III-B}, we set $K_0$ PSPs with DoD $1$. Consider  ${K}_0\geq 3$. It is then possible to estimate the UE's position using these $K_0$ SPs, which will be used as the initial UE's position, denoted by $\tilde{\mathbf{p}}$. Following the \emph{time difference of arrival} (TDoA)-based localization explained in Appendix A of the extended version, the position estimate $\tilde{\mathbf{p}}$ can be derived as
\begin{equation}
    \tilde{\mathbf{p}} = \left({\mathbf{A}}^{\top}{\mathbf{A}}\right)^{-1}{\mathbf{A}}^{\top}\left(d_{k_\text{ref}}{\mathbf{c}}+{\mathbf{b}}\right),
    \label{eq: initial localization}
\end{equation}
where $d_{k_\text{ref}}=c\tau_{k_\text{ref}}$ is the distance between $k_\text{ref}$-th RIS tile and UE which have the smallest ToA $\tau_{k_\text{ref}}$. The matrix $\mathbf{A}$, the vectors $\mathbf{c}$ and $\mathbf{b}$ are given as
\begin{align}
    {\mathbf{A}} &= \left[\mathbf{p}_{k(L)}-\mathbf{p}_{k_\text{ref}},\ldots, \mathbf{p}_{k(L-K_0+1)}-\mathbf{p}_{k_\text{ref}}\right]^{\top} \in\mathbb{R}^{K_0 \times 3},\nonumber\\
    {\mathbf{c}} &= \left[-\Gamma_{k(L)},\ldots, -\Gamma_{{k(L-K_0+1)}}\right]^{\top}\in\mathbb{R}^{K_0},\nonumber\\
    {\mathbf{b}} &= \frac{1}{2}
    \begin{bmatrix}
    \lVert\mathbf{p}_{k(L)}\rVert^2- \lVert\mathbf{p}_{k_\text{ref}}\rVert^2-\Gamma_{k(L)}^{2}\\ \vdots \\ \lVert\mathbf{p}_{k(L-K_0+1)}\rVert^2- \lVert\mathbf{p}_{k_\text{ref}}\rVert^2-\Gamma_{k(L-K_0+1)}^{2}
    \end{bmatrix}
    \in\mathbb{R}^{K_0},\nonumber
\end{align}
where $\Gamma_{k(i)}=(\tau_{k(i)}-\tau_{k_\text{ref}})c-\left\|\mathbf{p}_{k_\text{ref}} -\mathbf{p}_{k(i)} \right\|$ with the position of $k_\text{ref}$-th RIS tile. Fig. \ref{Fig: Initialization} graphically illustrates the SPL process with DoD~$1$.

\subsubsection{SPL with Geometric Discriminant (DoD is two)}
Consider a PSP whose DoDs are two, i.e., $|\mathcal{L}(i)|=2$. The sets  
\begin{align}
\mathcal{L}{(i)}=
\{k_1{(i)}, k_2{(i)}\}, \quad \mathcal{T}{(i)}=\{\tau^{p}{(i)}, \tau^{q}{(i)}\}.  \label{Set_DoD2}
\end{align}
We arrange the ToAs in $\mathcal{T}^{(i)}$ in descending order of ToA, i.e., $\tau^{p}{(i)} \ge \tau^{q}(i)$. Given the initial position estimate $\tilde{\mathbf{p}}$ of \eqref{eq: initial localization}, we operate SPL and update the position estimate alternatively. For example, by plugging $\tilde{\mathbf{p}}$ into the geometric discriminant in Theorem \ref{Theorem2}, we derive the discriminant region $\mathcal{R}{(i)}$ as
\begin{align}
    \mathcal{R}{(i)}= \left\{\mathbf{p}\in \mathbb{R}^2| f_{k_{1}{(i)},k_{2}{(i)}}(x,y)\ge 0,\right.\nonumber\\
    \left.x\ge \Upsilon_x{(i)}, y<\Upsilon_y{(i)}\right\},
    \label{eq: i-th region}
\end{align}
where $f_{k_{1}{(i)},k_{2}{(i)}}(x,y)$, $\Upsilon_x{(i)}$, and $\Upsilon_y{(i)}$ can be derived by plugging the coordinates of the RIS elements $k_{1}{(i)}$ and $k_{2}{(i)}$ into \eqref{eq: hyperbolic function}.  Then, we use $\tilde{\mathbf{p}}$ for the SPL of $\mathcal{T}{(i)}$, namely, 
\begin{equation}
    \bigg(\phi\big(\tau^{p}{(i)}\big),\phi \big(\tau^{q}{(i)}\big)\bigg)=
    \begin{cases}
    \big(k_1{(i)},k_2{(i)}\big), & \text{if } \tilde{\mathbf{p}}\in \mathcal{R}{(i)},\\
    \big(k_2{(i)},k_1{(i)}\big), & \text{otherwise.}
    \end{cases}
    \label{eq:i-th labeling}
\end{equation}
Last, we replace the current position estimate with the new one using   \eqref{eq: initial localization} after updating the matrix and vectors therein, given as  
\begin{align}
    &\mathbf{A} \leftarrow
    \left[\mathbf{A}; \mathbf{p}_{\phi(\tau^{p}(i))}-\mathbf{p}_{k_\text{ref}};\mathbf{p}_{\phi(\tau^{q}(i))}-\mathbf{p}_{k_\text{ref}}\right]^{\top}, 
    \nonumber\\
    &\mathbf{c} \leftarrow \left[\mathbf{c},-\Gamma_{\phi(\tau^{p}(i))},-\Gamma_{\phi(\tau^{q}(i))}\right]^{\top},
    \nonumber\\
    &\mathbf{b} \leftarrow \frac{1}{2}
    \begin{bmatrix}
    \mathbf{b}\\
\lVert\mathbf{p}_{\phi(\tau^{p}(i))}\rVert^2- \lVert\mathbf{p}_{k_\text{ref}}\rVert^2-\Gamma_{\phi(\tau^{p}(i))}^{2}\\  \lVert\mathbf{p}_{\phi(\tau^{q}(i))}\rVert^2- \lVert\mathbf{p}_{k_\text{ref}}\rVert^2-\Gamma_{\phi(\tau^{q}(i))}^{2}
    \end{bmatrix}.
    \nonumber
\end{align}
Fig. \ref{Fig: Greedy} graphically illustrates the SPL process with DoD~$2$.

\begin{algorithm}[t]
        \caption{Signal Path Labeling}\label{Geometric Signal Path Labeling}
        \begin{algorithmic}[1]
            \Statex Inputs: $\{\mathcal{L}(i)\}$, $\{\mathcal{T}{(i)}\}$, $\{\beta_k\}$, $\{\mathbf{p}_{k}\}$
            \Statex Output: The UE's position estimate $\tilde{\mathbf{p}}$
            \State \textbf{SPL with exclusive PSPs:}
            \State Collect ToAs of $K_0$ SPs whose DoD are one and label them according to  \eqref{eq:initial labeling}.
            \State Compute the UE's position estimate 
            $\tilde{\mathbf{p}}$ using \eqref{eq: initial localization}.
            \State \textbf{SPL with geometric discriminant:}
            \For{$M=2,3,\cdots$}
            \State Collect ToAs of SPs in $\mathcal{T}{(i)}$ and the corresponding RIS elements in $\mathcal{L}(i)$ whose DoD are $M$
            \If {$M=2$}
            \State Compute the discriminant region $\mathcal{R}(i)$ of \eqref{eq: i-th region} using the latest position estimate $\tilde{\mathbf{p}}$. 
                \State Label the SPs' ToAs according to \eqref{eq:i-th labeling}.
                \Else
                \State Go to \textbf{Algorithm \ref{Algorithm 2}}.
            \EndIf
            \State Update the UE's position $\tilde{\mathbf{p}}$. 
            \EndFor
            \State return $\tilde{\mathbf{p}}$.
        \end{algorithmic}
    \end{algorithm}



\subsubsection{SPL based on Geometric Discriminant (DoD is no less than three)}

Consider a PSP whose DoD is no less than three, i.e., $|\mathcal{L}{(i)}|\geq 3$, whose sets are given as \eqref{eq: L} and \eqref{eq: T}.  
The main idea is to rearrange the RIS elements in $\mathcal{L}(i)$ through pairwise geometric discriminant processes.
Take an example from the following initial hypothesis: 
\begin{align}
&\left(\phi(\tau^{p_{1}}{(i)}), \cdots, \phi(\tau^{p_{j}}{(i)}), \phi(\tau^{p_{j+1}}{(i)}),\cdots, \phi(\tau^{p_{|\mathcal{L}{(i)}|}}{(i)})\right)\nonumber\\
&=\left(k_1{(i)}, \cdots, k_j{(i)}, k_{j+1}{(i)},\cdots, k_{|\mathcal{L}{(i)}|}{(i)}\right).
\label{eq: hypothesis}
\end{align}
Using the current position estimate $\tilde{\mathbf{p}}$, we check that the adjacent SPs' labels $k_j{(i)}$ and $k_{j+1}{(i)}$ are correct through the geometric discriminant. If yes, we move on to verifying the subsequent adjacent SPs, say $k_{j+1}{(i)}$ and $k_{j+2}{(i)}$. 
Otherwise, we swap their labels, namely,
\begin{align}
\phi(\tau^{p_{j}}{(i)})=k_{j+1}{(i)}, \quad \phi(\tau^{p_{j+1}}{(i)})=k_{j}{(i)}. 
\end{align}
The process repeats until all adjacent pairs in the revised hypothesis are correct. The process is similar to the well-known bubble sort algorithm and thus called \emph{sorting} (SR)-based SPL algorithm. It is not dependable on the initial hypothesis.

Note that the SR-based SPL algorithm relies on the geometric discriminant between adjacent RIS elements, while those between other pairs are not made. Because we use the estimated position $\tilde{\mathbf{p}}$ instead of the ground truth, there is an inevitable error in the position estimate. 
It sometimes causes a logical error that the geometric discriminant between the SPs $p_j{(i)}$ and $p_{j+2}{(i)}$ fails though those between $p_j{(i)}$ and $p_{j+1}{(i)}$, and $p_{j+1}{(i)}$ and $p_{j+2}{(i)}$ are verified. Mathematically, the discriminant region $\mathcal{R}{(i)}$ can be expressed as the intersection of each label pair's region, namely,
\begin{align}
\mathcal{R}{(i)}=\bigcap_{m=1}^M\bigcap_{j=m+2}^M \mathcal{R}_{\phi(\tau^{m}{(i)}),\phi(\tau^{j}{(i)})},
\label{eq: region(general)}
\end{align}
where $\mathcal{R}_{\phi(\tau^{m}{(i)}),\phi(\tau^{j}{(i)})}$ is the discriminant region between the RIS elements $\phi(\tau^{m}{(i)})$ and $\phi(\tau^{j}{(i)})$. We skip the regions between adjacent elements, which have already been checked in the SR-based SPL. As a result, the resulting label sequence after the SR-based SPL can be said to have a logical error if $\tilde{\mathbf{p}}\notin \mathcal{R}^{(i)}$.

\begin{algorithm}[t]
        \caption{SPL when DoD is more than three}\label{Algorithm 2}
        \begin{algorithmic}[1]
            \Statex Inputs: $\{\mathcal{L}(i)\}$, $\{\mathcal{T}{(i)}\}$, $\{\beta_k\}$, $\{\mathbf{p}_{k}\}$, $\mathbf{p}_{\text{BS}}$, $M$
            \Statex Output: $\left(\phi(\tau^{p_{1}}{(i)}), \cdots, \phi(\tau^{p_{M}}{(i)})\right)$
\State Set the initial sequence:
\Statex \qquad $(p_{1}(i), \dots, p_{M}(i))=(k_1{(i)},\dots,k_M{(i)})$
\For{$m = 1$ \textbf{to} $M$}
\For{$j = m$ \textbf{to} $M-1$}
\State Derive the discriminant region $\mathcal{R}(i)$ between RIS elements $p_j(i)$ and $p_{j+1}(i)$ using \eqref{eq: Region}.  
\If{$\tilde{\mathbf{p}}\notin \mathcal{R}(i)$}
\State $\text{temp}=p_j(i)$
\State $p_j(i)=p_{j+1}(i)$
\State $p_{j+1}(i)=\text{temp}$
\EndIf
\EndFor
\EndFor

\State $\left(\phi(\tau^{p_{1}}{(i)}), \cdots, \phi(\tau^{p_{M}}{(i)})\right)=(p_{1}(i), \cdots, p_{M}(i))$.
\State Derive the discriminant region $\mathcal{R}(i)$ by intersecting the regions of non-adjacent label pairs using \eqref{eq: region(general)}.
\If{$\tilde{\mathbf{p}}\notin \mathcal{R}(i)$,}
\State Find the optimal label ${\mathcal{Q}{(i)}}^{*}$ to minimize the RE according to \eqref{eq: Residual error min}.
\EndIf
\end{algorithmic}
\end{algorithm}

We can overcome the above logical error by comparing every possible label sequence's \emph{residual error} (RE), defined as the aggregate absolute difference between measured and estimated values. Specifically, given a candidate label sequence $\mathcal{Q}{(i)}=(p_{1}(i),\ldots,p_{M}(i))$, the RE can be computed as
\begin{equation}
e(\mathcal{Q}{(i)})=\sum_{m=1}^{M}\left\lvert \Gamma_{p_{m}(i)}- \hat{\Gamma}_{p_{m}(i)}\right\rvert,
\label{eq: Residual error}
\end{equation}
where $\Gamma_{p_{m}(i)}$ is the measured distance difference between the $p_{m}(i)$-th SP and the LoS path, given as $\Gamma_{p_{m}(i)}=(\tau_{p_{m}(i)}-\tau_{0})c$.  On the other hand, $\hat{\Gamma}_{p_{m}(i)}$ represents the estimated distance difference using the position estimate $\tilde{\mathbf{p}}$, given as
\begin{align}
\hat{\Gamma}_{p_{m}(i)} = \left(\lVert \tilde{\mathbf{p}}-\mathbf{p}_{p_{m}(i)}\rVert+\lVert \mathbf{p}_{\text{BS}}-\mathbf{p}_{p_{m}(i)} \rVert\right)-\lVert \tilde{\mathbf{p}}-\mathbf{p}_{k_\text{ref}}\rVert.
\nonumber
\label{eq: estimated distance difference}
\end{align}
Here, the first term is the distance from the BS to the UE via the $q_m$-th RIS element, and the second is the direct distance.

Then, the sequence of the optimal label, denoted by ${\mathcal{Q}{(i)}}^*$, is determined by minimizing the RE, given as
\begin{equation}
{\mathcal{Q}{(i)}}^*=\argmin e(\mathcal{Q}{(i)}).
\label{eq: Residual error min}
\end{equation}
Note that the number of possible label sequences is $\mathcal{O}(M!)$, whose computation complexity is higher than the SR-based one. We only use the RE-based SPL for computational efficiency unless the sorting-based one works. The detailed algorithm is elaborated on Algorithm \ref{Algorithm 2}. 

\section{Computational Complexity Analysis}
\label{Sec:V}
The complexity is mainly determined by the IDFTs required to obtain the response in the time domain for SP decomposition, the sorting method used to label SPs, and the least square used to estimate the position. More details are explained as below:
\begin{enumerate}
    \item \textbf{Signal Path Decomposition}: The complexity of SPD is mainly determined by the IDFTs required to obtain the response in the time domain for ToA estimation. By operating 2D-IDFT in \eqref{eq: general 2D-IDFT}, we construct a 2D map spectrum of $N \times L$ size with $N$ subcarriers and $L$ PSPs, so the computational complexity is $\mathcal{O}(N L \log_2(N L))$. Unlike a 2D-IDFT, 1D grid search only performs IDFT in one dimension of size $N$, so the computational complexity is $\mathcal{O}( N  \log_2( N))$.
    \item \textbf{Signal Path Labeling}: The complexity of SPL is mainly determined by the sorting method to label SPs and TDoA localization. In the sorting method, the process repeats until all adjacent pairs in the revised hypothesis are correct. The process is similar to the well-known bubble sort algorithm, so its computation complexity is equivalent to that of the bubble sort algorithm as $\mathcal{O}(M^2)$, where $M$ represents the DoD of the concerned PSP. Note that the computation of the pair decision is not required since the pairing check is a one-shot decision based on the geometric discriminant of \eqref{eq:labeling}. Note that $M$ is determined by $N$ and $L$, as $\lfloor\frac{K}{L}\rfloor$, the computation is represented as $\mathcal{O}(\frac{K^2}{L^2})$. The complexity of the TDoA localization involves the inversion of a matrix of dimension $K \times K$, so the computational complexity is $\mathcal{O}(K^3)$. The whole procedure of SPL is performed repeatedly for all $M$, which is $L$ times, so the complexity is $\mathcal{O}(\frac{K^2}{L}+LK^3)$. 
\end{enumerate}
The overall computational complexity of the proposed 2D-SPC is the order of $\mathcal{O}(N L \log_2(N L))+\mathcal{O}(\frac{K^2}{L}+LK^3)$ where each term sequentially accounts for the complexity of SPD and SPL with TDoA localization.
\begin{remark}
    (2D-SPC vs. Direct localization) The complexity of the direct localization is $\mathcal{O}(NLG)$ where $G$ represents the number of test points. If $M$ is larger than $\log_2(NL)$, the direct localization is more complex than the proposed 2D-SPC. Given the simulation setting of this paper (i.e., $K=64$, $L=64$, $N=3200$), $\log_2(NL)$ is smaller than 18, and it is not a sufficient number for ML operations. Therefore, it can be seen that the proposed 2D-SPC has lower computational complexity compared to direct localization in practical scenarios.
\end{remark}

\begin{figure}[t]
\centering
\includegraphics[width=0.94\linewidth]{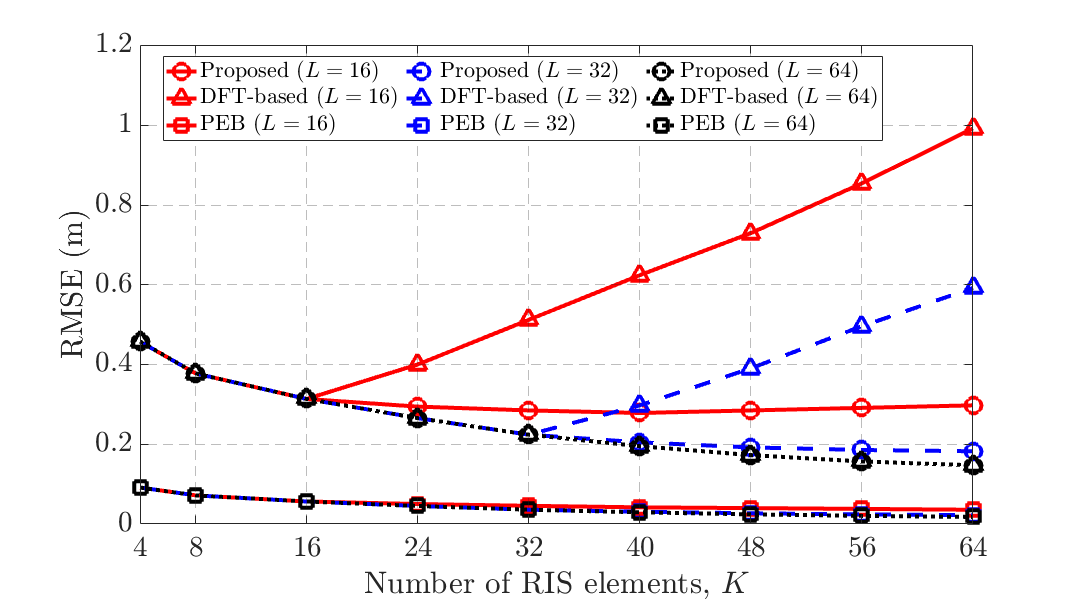}
\caption{Localization RMSE and PEB versus the number of RIS elements $K$. \label{Fig: RMSE_K}}
\end{figure}

\begin{figure}[t]
\centering
\includegraphics[width=0.94\linewidth]{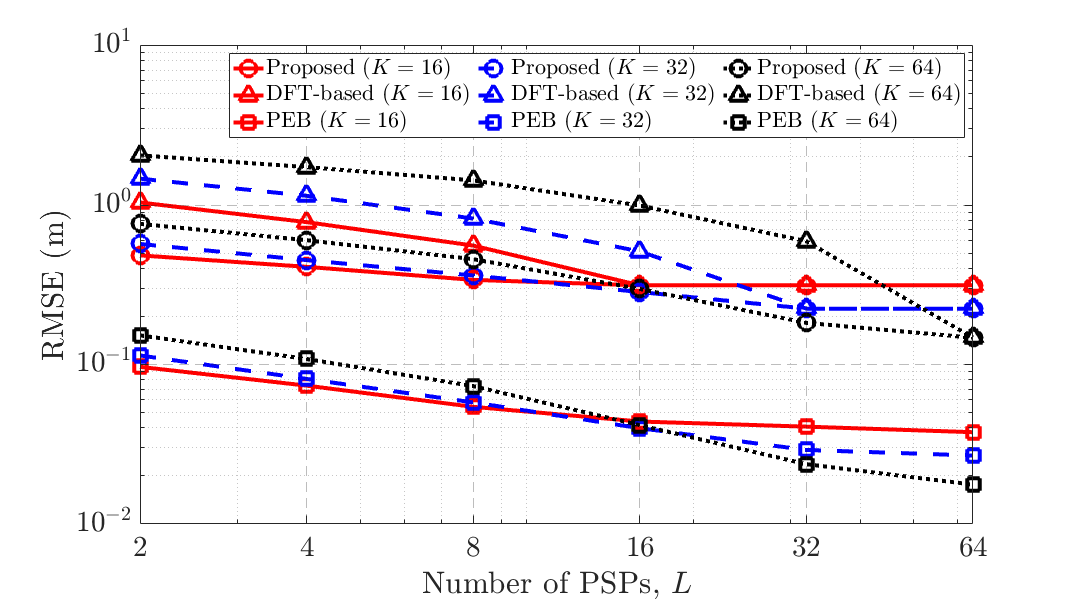}
\caption{Localization RMSE and PEB versus the number of PSPs $L$. \label{Fig: RMSE_L}}
\end{figure}
\begin{figure}[t]
\centering
\includegraphics[width=0.94\linewidth]{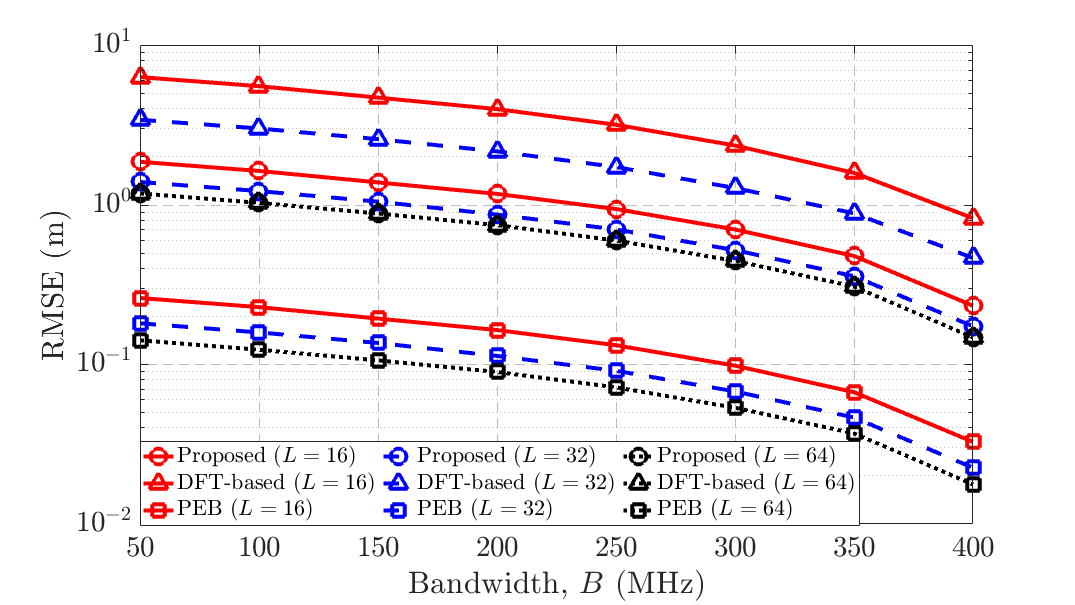}
\caption{Localization RMSE and PEB versus bandwidth $B$. \label{Fig: RMSE_B}}
\end{figure}

\begin{figure*}
\centering
\begin{subfigure}{0.31\textwidth}
    \includegraphics[width=\textwidth]{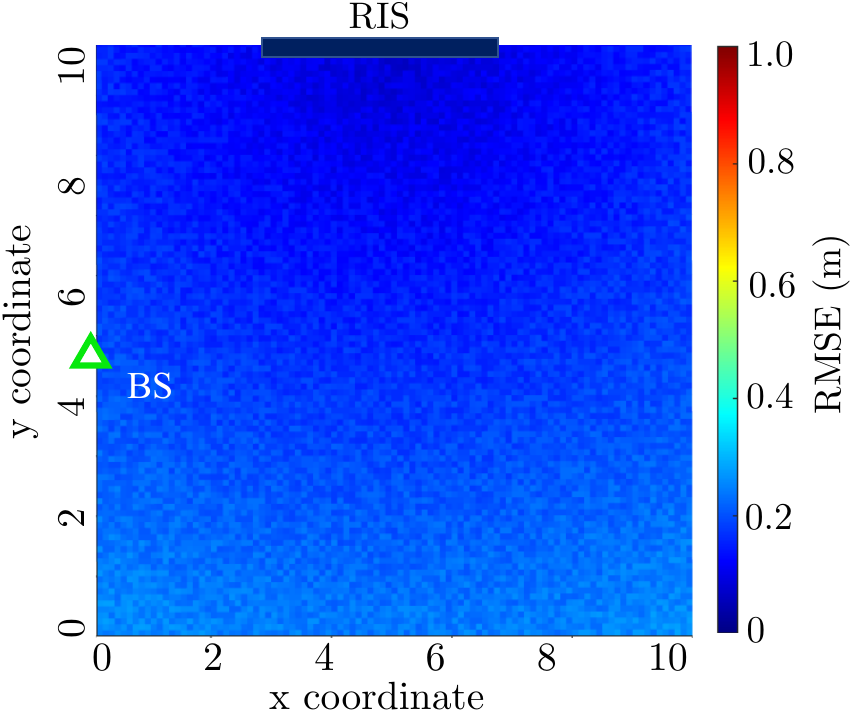}
    \caption{Proposed ($L=32$)}
    \label{fig:HM-1}
\end{subfigure}
\hfill
\begin{subfigure}{0.31\textwidth}
    \includegraphics[width=\textwidth]{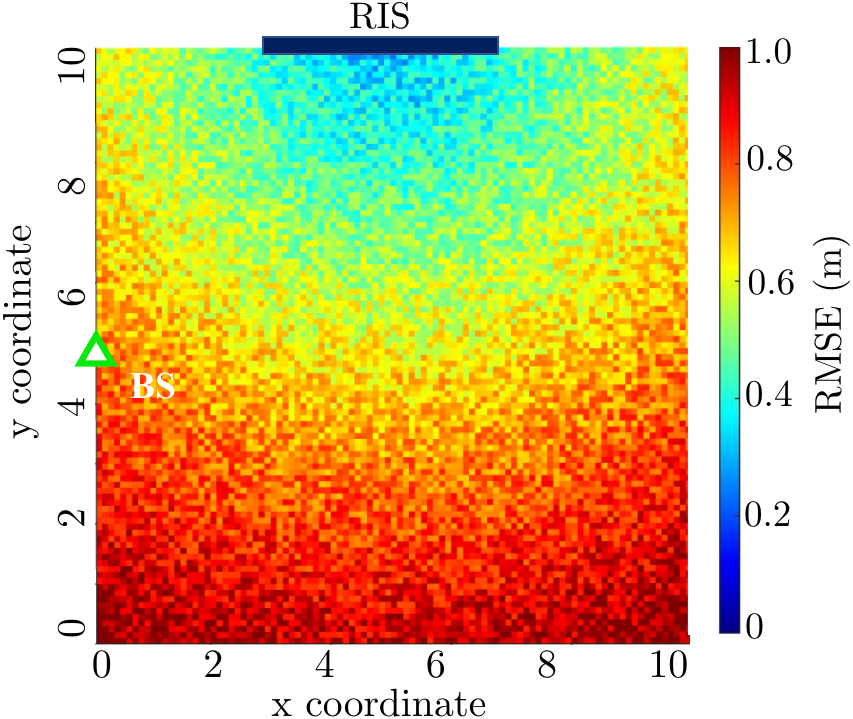}
    \caption{DFT-based ($L=32$)}
    \label{fig:HM-2}
\end{subfigure}
\hfill
\begin{subfigure}{0.31\textwidth}
    \includegraphics[width=\textwidth]{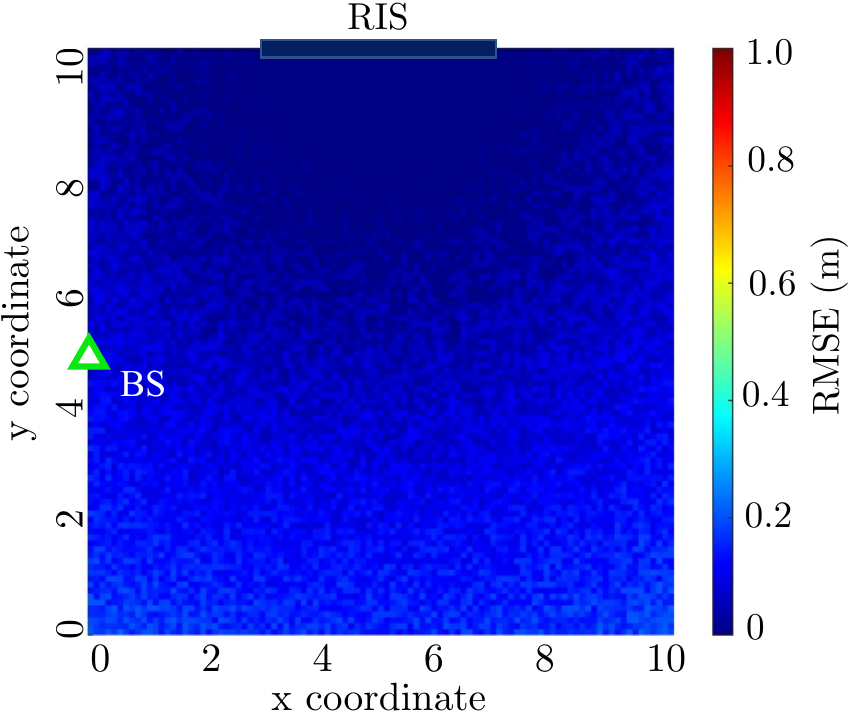}
    \caption{DFT-based ($L=64$)}
    \label{fig:HM-3}
\end{subfigure}
        
\caption{RMSE heatmap when the UE is placed in the area $(0,0)\times(10,10)$  $\text{m}^2$.}
\label{fig:HM}
\end{figure*}

\begin{figure}[t]
\centering
\includegraphics[width=0.94\linewidth]{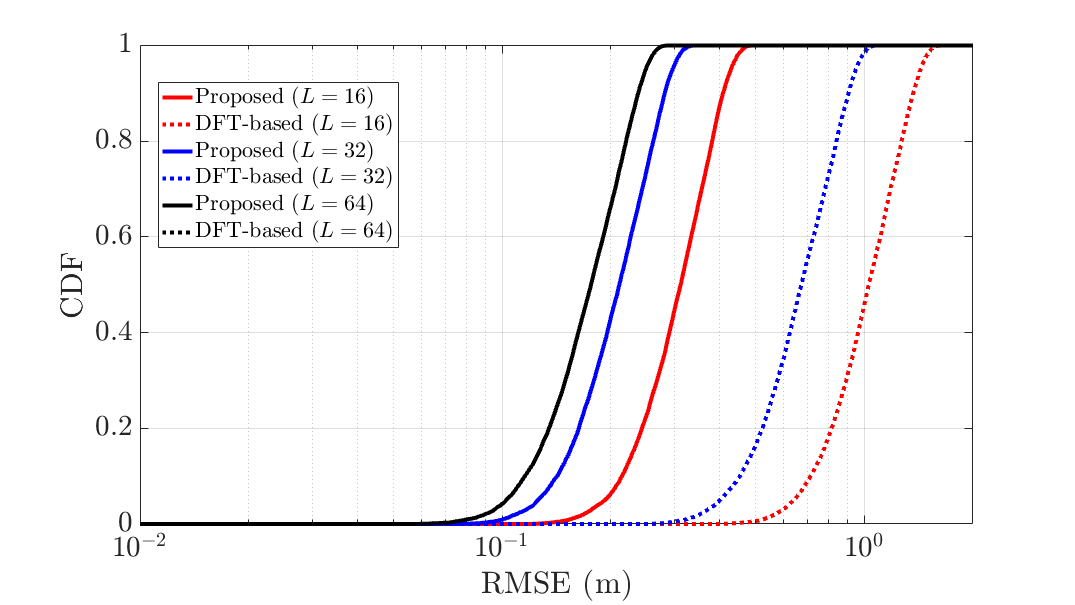}
\caption{CDF vs. RMSE when the UE is placed in the area $(0,0)\times(10,10)$  $\text{m}^2$. \label{Fig: CDF_RMS}}
\end{figure}
\section{Simulation Results}
\label{Sec:VI}

This section presents some numerical results using the simulation parameters summarized below unless specified. The size of the indoor room is $10\times10\times3$ ($\text{m}^3$). As illustrated in Fig. \ref{Fig: Geometric Model}, the concerned scenario has a linear RIS of $K=64$, with element spacing $d_{\text{RIS}}=0.1$ (m) and $M=40$ $(M_x \times M_z=4 \times 10)$. The RIS is distributed in a line ($y=10$) with the center $[5, 10, 2]^{\top}$ (m). The BS is located at $[0,5,2]^{\top}$ (m), whereas the UE is randomly located with uniform distribution in the area $(0,0,0)\times(10,10,0)$ $\text{m}^3$.
The OFDM signal consists of $N$ sub-carriers with frequency spacing $\delta=120$ kHz at the center frequency of $28$ GHz. Number of the sub-carriers $N$ is set to $3200$, and the $B$ is approximated to $400$ MHz ($B=3200 \times 120 \text{ kHz} \approx 400\text{ MHz}$). The oversampling factor $Q_c$ is set to 4. The total transmission power $P$ is set to $20$ dBm, and the UE noise power $N_0$ is $-8$ dBm. An asynchronous system is considered in which we model the clock and phase offsets $t_0$ and $\varphi_0$ as random variables uniformly distributed in $[0, T_a]$
and $[0, 2\pi)$, respectively, where $T_a=10^{-6}$ sec denotes the clock mismatch
uncertainty. For multipath, simulation settings are set the same as \cite{9625826}. For a benchmark, the DFT-based algorithm in \cite{9625826} is considered where the phase shift of each element follows a different row in a DFT matrix. The theoretical lower bound of UE’s position estimation, referred to as position error bound (PEB), is also provided as a benchmark (Derivation of PEB is explained in Appendix D of the extended version).

We first examine the localization performance of the proposed SPC algorithm. Fig. \ref{Fig: RMSE_K} shows the localization \emph{root mean square error} (RMSE) averaged over $1000$ Monte Carlo iterations and PEB as the number of RIS elements $K$ varies. Under the loose latency condition ($L\ge K$), it is observed that both proposed and DFT-based algorithms have the same localization performance since all SPs are labeled with a one-to-one mapping to each PSP. Conversely, in the strict latency condition ($L<K$), it is observed that the proposed algorithm provides better localization accuracy than DFT-based. The RMSE of DFT-based increases since SPs that have equivalent PSP can not be labeled correctly. Compared to DFT-based, the RMSE of the proposed one slightly decreases since SPs with equivalent PSP are labeled well by geometric discriminant explained in Sec. \ref{Sec:IV}. Since the PEB is proportional to the accuracy of ToA estimates, the PEB also decreases as $K$ increases.


\begin{figure}[t]
\centering
\includegraphics[width=0.94\linewidth]{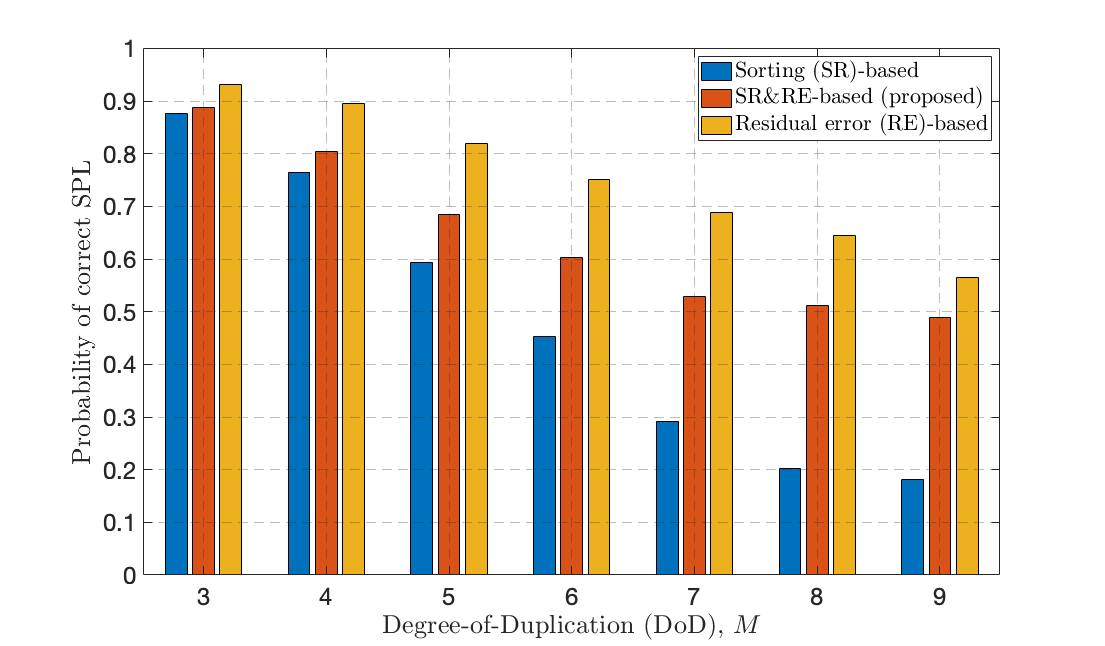}
\caption{Probability of correct SPL vs. degree-of-duplication (DoD) $M$. \label{Fig: P_SPL_M}}
\end{figure}

Fig. \ref{Fig: RMSE_L} shows the localization RMSE averaged over $1000$ Monte Carlo iterations and PEB the number of PSPs $L$ varies. Same as Fig. \ref{Fig: RMSE_K}, both proposed and DFT-based algorithms have the same localization performance in the loose latency condition ($L \geq K$). In the case of $L \geq K$, the localization performance is constant even if $L$ increases because extra PSPs exceeding $K$ are redundant and not used. As $L$ increases, the RMSE of proposed and DFT-based decreases due to the number of RIS elements with an equivalent PSP $M$ decreasing as $M \approx (K-4)/(L-4)$ Similarly to Fig. 7, the PEB also decreases as $L$ increases since the PEB is proportional to the accuracy of ToA estimates.


The localization RMSE averaged over $1000$ Monte Carlo iterations the bandwidth $B$ varies in Fig. \ref{Fig: RMSE_B}. As expected, the proposed algorithm has better accuracy than the DFT-based regardless of bandwidth $B$. Especially when $B=50$ MHz and $L=16$, the localization RMSE of 2D-SPC is $1.85$ m, while that of DFT-based one is $6.3$ m, verifying the discussion in Remark 1.
In case of $B=400$ MHz, the localization RMSE are $0.28$ m (proposed, $L=16$), $1.02$ m (DFT-based, $L=16$) and $0.17$ m (both, $L=64$), respectively. As $B$ decreases, the PEB drops significantly because the theoretical TOA resolution is limited by the inverse of $B$, as indicated by the analysis in Sec. III. PEB are $0.24$ m ($B=50$ MHz) and $0.08$ m ($B=400$ MHz), respectively.
It is confirmed that the proposed algorithm can estimate more precisely than the resolution over the bandwidth since there is no issue of overlapping SPs.

Fig. \ref{fig:HM} shows the localization RMSE heatmap of the proposed and DFT-based SPCs and how the localization error is distributed within the concerned area of $(0,0)\times(10,10)$ ($\text{m}^2$). 
It is observed that the localization RMSE in the area is within $1.2$ m (proposed, $L=32$), $3.0$ m (DFT-based, $L=32$), $1.0$ m (DFT-based, $L=64$), respectively. As expected, the DFT-based algorithm ($L=32$) has a poor localization performance, whereas the proposed algorithm ($L=32$) has a similar ~localization performance to the DFT-based algorithm ($L=64$). As the UE moves away from the RIS, the localization performance decreases for the following two reasons. The first is higher path loss attenuation, negatively affecting SPC and ToA estimations. Second, a longer distance to the RIS causes the change of the propagation condition from NF to FF, hampering the geometric localization.

The corresponding analysis in the \emph{cumulative distribution function} (CDF) of RMSE is represented in Fig. \ref{Fig: CDF_RMS}. The DFT-based algorithm ($L=64$) is the lower bound where both 90\% and 50\% percentile error are $0.21$ m and $0.18$ m, respectively. The 90\% percentile errors are $0.43$ m (proposed, $L=16$), $0.21$ m (proposed, $L=32$), $1.76$ m (DFT-based, $L=16$), and $0.64$ m (DFT-based, $L=32$) while 50\% percentile error are $0.34$ m, $0.18$ m, $1.09$ m, and $0.53$ m, respectively. Conversely, a localization accuracy better than $1$ m is obtained in more than 100\% (proposed, $L=32$) and 100\% (proposed, $L=16$) compared to 87\% (DFT-based, $L=32$) and 32\% (DFT-based, $L=16$). It means the proposed algorithm is suitable for indoor localization even if strict latency conditions due to the constraint of discrete phase shift \cite{9424177}.


Finally, in Fig. \ref{Fig: P_SPL_M}, the probability of correct SPL is investigated for SR-based, RE-based, and SR\&RE-based SPL (proposed).
The proposed SPL is a hybrid method that performs SR-based first, then switches to RE-based when there is a logical error. Conversely, both SR-based and RE-based SPLs as benchmarks are standalone. As $M$ increases, the probability of correct SPL decreases since the number of SPL combinations is factorially increased by $M!$.
The performance of GR-based drops drastically compared to RE-based.
As $M$ increases, the labeling intersection of GR-based can be empty set easily, which can cause a false alarm of SPL because of the error between the ground truth and the estimated location. Conversely, the RE-based computes the residual error of all possible SPL candidates and then chooses the best candidate, minimizing residual error.
There is a trade-off between the computational complexity and SPL performance. The GR\&RE-based (proposed) switch from GR-based to RE-based, so it is similar to GR-based (when $M$ is small) or RE-based (when $M$ is large).
Therefore, the proposed makes a good balance between performance and computational complexity regardless of $M$.

\section{Conclusions}
\label{Sec:VII}
In this paper, we have proposed a 2D-SPC comprising SPD and SPL to unleash the full degree of freedom of RIS-assisted NF localization. SPD can be efficiently designed in a 2D spectrum map constructed by the RIS’s PSP and OFDM transmissions, providing an additional dimension to resolve SPs. Besides, 2D-SPC enables the UE to achieve precise localization even in insufficient PSPs through a novel SPL algorithm based on the discriminant derived from the UE and RIS geometric relation.
The proposed 2D-SPC's effectiveness is verified through analytic and numerical studies, showing that it outperforms the benchmark in various network settings.
For further study, incorporating angle with ToA can expand a new dimension for SPC, and it can be solved by multi-dimension SPC such as Tensor decomposition.

\begin{appendices}
\section{TDoA Based Localization}
\label{app:AppA}
Assume that all SPs $p$ are perfectly labeled to ground truth. Given $\{\tau_k\}$ of \eqref{eq: time delay}, we can localize the UE's position $\mathbf{p}=[x,y,z]^{\top}$ by following the TDoA-based algorithm \cite{1458275}. Denote $d_k$ the distance between the UE and the $k$-th RIS tile, say $d_k=\left\|\mathbf{p} -\mathbf{p}_{k} \right\|$.
From \eqref{delay}, $d_k$ can be written in terms of $\tau_k$ as
\begin{align}
    d_{k} =
    c\left(\tau_k-t_0 \right)-{\left\|\mathbf{p}_{\text{BS}} -\mathbf{p}_{k} \right\|}, \quad k=1,\ldots,K, 
    \label{eq:estimated distance(Async)}
\end{align}
where $k_{\text{ref}}$ is the reference RIS tile that has a minimum ToA, and the BS-UE constant synchronization gap $t_0$ is specified in \eqref{delay}. To cancel out the effect of $t_0$, we derive the distance difference compared to the LoS path $\Gamma_k=d_k-d_{k_\text{ref}}$ as\footnote{When LoS path exists, the distance difference can be measured compared with the LoS path.}
\begin{align}
    \Gamma_{k}  
    &=(\tau_{k}-t_0)c-(\tau_{k_\text{ref}}-t_0)c-\left\|\mathbf{p}_{k_\text{ref}} -\mathbf{p}_{k} \right\|\nonumber\\ 
    &=\Delta\tau_k c-\left\|\mathbf{p}_{k_\text{ref}} -\mathbf{p}_{k} \right\|,\quad k=1,\ldots,K,
    \label{eq:TDoA(Async)}
\end{align}
where the term $\Delta\tau_k=\tau_k-\tau_{k_\text{ref}}$ represents the signal path $k$'s TDoA. 

Next, following from a basic geometry between two points in a 3D coordinate system, the distance $d_k$ can be differently expressed in terms of components $x$ and $y$, given as
\begin{align}
    d_k=
    \sqrt{(x_k-x)^2 + (y_k-y)^2 + (z_k-z)^2}, \quad k=1, \ldots, K.
    \label{eq:distance}
\end{align}
Then, we can derive the following equation:
\begin{align}\label{eq:diff_square}
    d_k^2-d_{k_\text{ref}}^2=&\underbrace{(x_k)^2 + (y_k)^2 + (z_k)^2 }_{={\lVert\mathbf{p}_{k}\rVert}^{2}}-\underbrace{(x_{k_\text{ref}})^2 + (y_{k_\text{ref}})^2 + (z_{k_\text{ref}})^2}_{={\lVert\mathbf{p}_{k_\text{ref}}\rVert}^{2}}\nonumber\\
    &-2\Big\{\underbrace{(x_k-x_{k_\text{ref}})x + (y_k-y_{k_\text{ref}}) y + (z_k-z_{k_\text{ref}}) z}_{=\left(\mathbf{p}_{k}-\mathbf{p}_{k_\text{ref}}\right)^{\top} \mathbf{p}}\Big\}.
\end{align}
Recalling $d_k=d_{k_\text{ref}}+\Gamma_k$, 
the above equation \eqref{eq:diff_square} can be equivalently transformed into a vector form as  
\begin{equation}
-\left(\mathbf{p}_{k}-\mathbf{p}_{k_\text{ref}}\right)^{\top} \mathbf{p} = \Gamma_{k}d_{k_\text{ref}} + \frac{1}{2}\left(\Gamma_{k}^2-{\lVert\mathbf{p}_{k}\rVert}^{2}+{\lVert\mathbf{p}_{k_\text{ref}}\rVert}^{2}\right).
\label{eq: rearranging term(Async)}
\end{equation}
Eventually, stacking the above $(K-1)$ equations forms a system of  linear equations as
\begin{equation}
    \mathbf{A}\mathbf{p} = d_{k_\text{ref}}\mathbf{c}+\mathbf{b},
    \label{eq: ToA fusion method(Async)}
\end{equation}
where
\begin{align}
    \mathbf{A}= &[\mathbf{p}_{1}-\mathbf{p}_{k_\text{ref}},\ldots, \mathbf{p}_{k_\text{ref}-1}-\mathbf{p}_{k_\text{ref}},\nonumber\\ &\quad \mathbf{p}_{k_\text{ref}+1}-\mathbf{p}_{k_\text{ref}},\ldots,\mathbf{p}_{K}-\mathbf{p}_{k_\text{ref}}]^{\top} \in \mathbb{R}^{(K-1)\times 2},\nonumber\\
    \mathbf{c} =& \left[-\Gamma_{1},\ldots,-\Gamma_{k_{\text{ref}}-1},-\Gamma_{k_{\text{ref}}+1},\ldots,-\Gamma_{K}\right]^{\top} \in \mathbb{R}^{K-1}, \nonumber\\
    \mathbf{b} = & \frac{1}{2}
    \begin{bmatrix}
    \lVert\mathbf{p}_{1}\rVert^2- \lVert\mathbf{p}_{k_{\text{ref}}}\rVert^2-\Gamma_{1}^{2}\\ \vdots \\
    \lVert\mathbf{p}_{k_{\text{ref}}-1}\rVert^2- \lVert\mathbf{p}_{k_{\text{ref}}}\rVert^2-\Gamma_{k_{\text{ref}}-1}^{2}\\
    \lVert\mathbf{p}_{k_{\text{ref}}+1}\rVert^2- \lVert\mathbf{p}_{k_{\text{ref}}}\rVert^2-\Gamma_{k_{\text{ref}}+1}^{2}\\\vdots \\
    \lVert\mathbf{p}_{K}\rVert^2- \lVert\mathbf{p}_{k_{\text{ref}}}\rVert^2-\Gamma_{K}^{2}
    \end{bmatrix}\in \mathbb{R}^{K-1}.\nonumber
\end{align}
Lastly, the UE's position $\mathbf{p}$ can be estimated by solving the linear system of \eqref{eq: ToA fusion method(Async)}, given as
\begin{equation}
    \mathbf{p} = \left(\mathbf{A}^{\top}\mathbf{A}\right)^{-1}\mathbf{A}^{\top}\left(d_{k_{\text{ref}}}\mathbf{c}+\mathbf{b}\right),
    \label{eq: TDoA localization}
\end{equation}
which has a unique solution when the number of elements $K$ is no less than $4$. 

\section{The proof of Theorem \ref{Theorem1}} \label{app:AppB}
The range of the 2D sinc function's main lobe, defined as twice the distance to the nearest zero-crossing point from the peak, are ${1}/{B}$ and ${1}/{L}$ in the domain of $\tau$ and $\beta$, respectively. In other words, each sinc function's main lobe does not overlap if there is no other sinc function within the range, completing the proof.

\section{The proof of Theorem \ref{Theorem2}} \label{app:AppC}
Given unlabeled SPs $p_{1}$ and $p_{2}$ satisfying $\tau_{p_{1}} \ge \tau_{p_{2}}$, the two SPs can be labeled according to the region delimited by the hyperbolic function of \eqref{eq: hyperbolic function}. Based on $x=\Upsilon_x$, the hyperbolic function consists of the left $\mathcal{H}_1$ (in $x<\Upsilon_x$) and right curve $\mathcal{H}_2$ (in $x>\Upsilon_x$) which are defined as
\begin{align}
    \mathcal{H}_1&=\left\{\mathbf{p}\in \mathbb{R}^2|-(d_{k_1}-d_{k_2})=d_{{k_2}}^{\text{BS}}-d_{{k_1}}^{\text{BS}}\right\},\\
    \mathcal{H}_2&=\left\{\mathbf{p}\in \mathbb{R}^2|d_{k_1}-d_{k_2}=d_{{k_2}}^{\text{BS}}-d_{{k_1}}^{\text{BS}}\right\}.
\end{align}
Note that $d_{{k_2}}^{\text{BS}}-d_{{k_1}}^{\text{BS}}$ is always positive since the BS is located to the left of the RIS. When the UE is in $x<\Upsilon_x$, $d_{k_1}-d_{k_2}$ is always negative. It satisfies the complementary region of $\mathcal{R}$, which enables us to label the SPs $p_{1}$ and $p_{2}$ as $k_2$ and $k_1$. In other words, $\mathcal{H}_{1}$ does not affect any geometric discriminant for SPL.
Based on the right curve $\mathcal{H}_2$, the UE's region can be divided into two parts. 
\begin{align}
    \mathcal{R}_{1}&=\{\mathbf{p}\in \mathbb{R}^2|d_{k_1}-d_{k_2}<d_{{k_2}}^{\text{BS}}-d_{{k_1}}^{\text{BS}}\},\label{eq: C3}\\
    \mathcal{R}_{2}&=\{\mathbf{p}\in \mathbb{R}^2|d_{k_1}-d_{k_2}\ge d_{{k_2}}^{\text{BS}}-d_{{k_1}}^{\text{BS}}\}\label{eq: C4}.
\end{align}
The $\mathcal{R}_{2}$ is included in $\mathcal{R}$ of \eqref{eq: set of hyperbolic} and $\mathcal{R}_{1}$ is included in vice versa of $\mathcal{R}$. We complete the proof. 

\section{Cramer-Rao Bound Derivation} \label{app:AppD}
Let $k_{\text{ref}}$ be the reference RIS tile that has a minimum ToA. According to TDoA-based localization in Appendix A, the $(K-1)$ estimated TDoA measurements $\boldsymbol{\Delta\tau}=[\Delta\tau_1, \ldots, \Delta\tau_{k_{\text{ref}}-1}, \Delta\tau_{k_{\text{ref}}+1},\ldots,\Delta\tau_K]^{\top}\in \mathbb{R}^{K-1}$ are given. We compute the CRB on UE's position estimation error given TDoA measurements. For unbiased estimators, we neglect the terms of clock bias, i.e., $t_0$, $\varphi_0$, thus obtaining an optimistic bound.
By following \cite{9625826}, the log-likelihood function $f(\boldsymbol{\Delta\tau}|\mathbf{p})$ is defined as 
\begin{equation}
    f(\boldsymbol{\Delta\tau}|\mathbf{p}) \propto -\sum_{k\in\mathcal{K}\setminus \{k_{\text{ref}}\}}\frac{(\Delta\tau-\mu_k)^2}{2\sigma_k^2}.
    \label{eq: Log-likelihood function}
\end{equation}
where the true TDoAs $\boldsymbol{\mu}=[\mu_1, \ldots, \mu_{k_{\text{ref}}-1}, \mu_{k_{\text{ref}}+1},\ldots,\mu_K]^{\top}$ with TDoA of $k$-th element $\mu_k$ and the error variance of the TDoA estimates $\sigma_k^2$. It is given by the CRB defined as $\sigma_k^2={1}/{(8\pi^2 B^2\zeta_k)}$, where $\zeta_k=(1/\mathsf{SNR}_k+1/\mathsf{SNR}_{k_\text{ref}})^{-1}$ is the combined SNR of the two SPs involved in the TDoA estimate \cite{4802191}. Then, the Fisher information matrix (FIM) $\mathbf{J}(\mathbf{p})$ can be derived as
\begin{align}
    \mathbf{J}(\mathbf{p})&=\mathbb{E}\left[\bigtriangledown_{\mathbf{p}}^{\top} f(\boldsymbol{\Delta\tau}|\mathbf{p}) \bigtriangledown_{\mathbf{p}} f(\boldsymbol{\Delta\tau}|\mathbf{p})\right] \nonumber \\
        &=\sum_{k\in\mathcal{K}\setminus \{k_{\text{ref}}\}} \frac{1}{\sigma_k^2} \bigtriangledown_{\mathbf{p}}^{\top} \mu_k \bigtriangledown_{\mathbf{p}} \mu_k,
    \label{eq: FIM}
\end{align}
where $\bigtriangledown_{\mathbf{p}} \mu_k=[\frac{\partial \mu_k}{\partial x},\frac{\partial \mu_k}{\partial y},\frac{\partial \mu_k}{\partial z}]$ is the gradient of the TDoAs, with elements given by
\begin{align}
    \frac{\partial \mu_k}{\partial x}&=\frac{x-x_k}{c d_k}-\frac{x-x_{k_\text{ref}}}{c d_{k_\text{ref}}},\\
    \frac{\partial \mu_k}{\partial y}&=\frac{y-y_k}{c d_k}-\frac{y-y_{k_\text{ref}}}{c d_{k_\text{ref}}},\\
    \frac{\partial \mu_k}{\partial z}&=\frac{z-z_k}{c d_k}-\frac{z-z_{k_\text{ref}}}{c d_{k_\text{ref}}}.
\end{align}
where $(x_{k_\text{ref}}, y_{k_\text{ref}}, z_{k_\text{ref}})$ is 3D coordinates of reference RIS tile and $d_{k_\text{ref}}$ is distance between UE and reference RIS tile.
Finally, the PEB can be calculated as
\begin{equation}
  \text{PEB} = \sqrt{\text{tr}   \left[\mathbf{J}^{-1}(\mathbf{p})\right]  },
  \label{eq: PEB}
\end{equation}
where $\text{tr}(\cdot)$ denote the trace operator.

\end{appendices}

\bibliographystyle{IEEEtran}
\bibliography{library}
\end{document}